\documentclass[aps,pra,twocolumn]{revtex4}%
\usepackage{amsfonts}
\usepackage{amsmath}
\usepackage{amssymb}
\usepackage{graphicx}
\usepackage[pdftex,bookmarks]{hyperref}
\providecommand{\U}[1]{\protect\rule{.1in}{.1in}}
\def\be{\begin{equation}}
\def\ee{\end{equation}}
\def\bea{\begin{eqnarray}}
\def\eea{\end{eqnarray}}
\def\bma{\begin{mathletters}}
\def\ema{\end{mathletters}}

\def\0{\overline{0}}

\def\q0{\underline{0}}

\def\U{{ U}}

\def\one{\leavevmode\hbox{\small1\normalsize\kern-.33em1}}

\begin{document}
\title{Preparation of Decoherence Free Cluster States with Optical Superlattices}
\author{Liang Jiang,$^{1}$ Ana Maria Rey,$^{2,3}$ Oriol Romero-Isart,$^{4}$ Juan
Jos\'{e} Garc\'{\i}a-Ripoll,$^{5}$ Anna Sanpera,$^{4,6}$ Mikhail D.
Lukin$^{1,2}$}
\affiliation{$^{1}$ Department of Physics, Harvard University, Cambridge, MA 02138, USA}
\affiliation{$^{2}$ Institute for Theoretical Atomic, Molecular and Optical Physics,
Cambridge, MA 02138, USA}
\affiliation{$^{3}$ JILA and Department of Physics, University of Colorado, Boulder,
Colorado 80309-0440, USA}
\affiliation{$^{4}$ Departament de F{\'\i}sica, Universitat Aut\`onoma de Barcelona,
E-08193 Bellaterra, Catalonia, Spain.}
\affiliation{$^{5}$ Facultad de CC. F{\'\i}sicas, Universidad Complutense de Madrid, Ciudad
Universitaria s/n, Madrid, E-28040, Spain.}
\affiliation{$^{6}$ Instituci\'o Catalana de Recerca i Estudis Avan\c cats, E-08010,
Barcelona, Spain}


\begin{abstract}
We present a protocol to prepare decoherence free cluster states using
ultracold atoms loaded in a two dimensional superlattice. The superlattice
geometry leads to an array of $2\times2$ plaquettes, each of them holding four
spin-1/2 particles that can be used for encoding a single logical qubit in the
two-fold singlet subspace, insensitive to uniform magnetic field fluctuations
in any direction. Dynamical manipulation of the supperlattice yields distinct
inter and intra plaquette interactions and permits to realize one qubit and
two qubit gates with high fidelity, leading to the generation of universal
cluster states for measurement based quantum computation. Our proposal based
on inter and intra plaquette interactions also opens the path to study
polymerized Hamiltonians which support ground states describing arbitrary
quantum circuits.

\end{abstract}
\date{\today}
\maketitle

\section{Introduction}

Quantum technology, in particular quantum information processing and quantum
metrology, requires the precise preparation of quantum states that outperform
a given task better than any classical strategy. As shown in recent years, the
unprecedent control and precision provided by ultracold gases in optical
lattices makes these systems optimal candidates for such a technology.
Furthermore, since the dynamical control over the optical lattice parameters
permits the simultaneous coupling between nearest atom-lattice sites, these
systems are also increasingly used as quantum simulators to mimic distinct
complex condensed matter Hamiltonians\cite{Bloch08}.

The controlled generation of double well lattices, i.e. lattices whose unit
cells contains two sites, has opened the possibility to isolate and address
individually pairs of atoms, and hence to manipulate the interactions between
them. Seminal results are the demonstration of controlled exchange interaction
between pairs of neutral atoms in an optical lattice when the atoms are forced
to be in the same location \cite{Anderlini07}, and the demonstration of
superexchange interactions \cite{Trotzky08}, showing that the interactions
between atoms trapped in two adjacent sites of the optical lattice can be made
analogous to the interactions between atomic spins in magnetic materials.
While Ref.~\cite{Anderlini07} sets a basis to perform in a controlled manner
two qubit gates between neighboring atoms in the double well lattice,
Ref.~\cite{Trotzky08} opens a direct path towards the realization of
low-temperature quantum magnets and a variety of many-body spin models with
ultracold atoms.

The atomic interaction control achieved with optical lattices has also direct
applications to quantum computation. A particularly well suited approach that
exploits the innate massive parallelism of such systems to perform quantum
computation is the Measurement Based Quantum Computation (MBQC), where
information is processed by means of a sequences of measurements on a highly
entangled initial state. It requires the capability to create
\textit{universal cluster state}, that is, a multipartite quantum state able
to reproduce any entangled quantum state in 2D, and to perform local single
qubits operations. Using as qubits two internal states of atoms in a 2D
optical lattice, it is possible to create a highly entangled quantum state by
means of controlled collisions\cite{Mandel03}, which is indeed a prerequisite
for the generation of universal cluster state.

Although neutral atoms couple weakly to the environment and they have
relatively long coherence times compared with the time scale associated with
the achievable coupling strength, when atoms are brought to an entangled state
decoherence will rapidly destroy any quantum superposition of atoms. The
larger the entangled system is, the faster it will decohere. To fight against
decoherence one should prepare the atoms in quantum states that are robust
against external perturbations. For periodic arrays of double-wells
\cite{Tame07,Rey07,Vaucher08}, resilient encoding schemes using two internal
states (two Zeeman levels) of the atoms have been proposed. In these schemes,
each double-well traps two two-level particles to encode a logical qubit. The
logical space is spanned by the singlet and triplet states of the two spin 1/2
particle along the quantization axis (here denoted by z) \{$\left\vert
S\right\rangle =\frac{1}{\sqrt{2}}\left(  \left\vert \uparrow\right\rangle
\left\vert \downarrow\right\rangle -\left\vert \downarrow\right\rangle
\left\vert \uparrow\right\rangle \right)  $ and $\left\vert T_{0}\right\rangle
=\frac{1}{\sqrt{2}}\left(  \left\vert \uparrow\right\rangle \left\vert
\downarrow\right\rangle +\left\vert \downarrow\right\rangle \left\vert
\uparrow\right\rangle \right)  $\}. Since these states have a zero z-component
of the total spin, such encoding is insensitive to fluctuations of the
magnetic field along the quantization axis. In practice, such an encoding
scheme
is very well suited for robust controlled interactions along one, lets say
horizontal, direction. To create a universal 2D cluster states (cluster states
in 1D are not universal), interactions and hence entanglement between
neighboring atoms along both, the horizontal and the vertical directions,
should be performed in such a way that robustness is preserved. In the above
encoding, interactions along the vertical direction, will leave the subspace
of zero spin component
along the quantization axis, becoming very fragile in front of external
magnetic field fluctuations noise. In this paper we show how this limitation
can be overcome by using 2D optical superlattices.

In passing, let us point out that superimposing secondary optical lattices (or
"superlattices") on top of the primary ones to further modify the potential in
which the atoms are trapped permits in general to create polymerized lattices.
By polymerized lattices we mean lattices consisting of weakly coupled groups
of neighboring atomic sites denoted as plaquettes. An example of a polymerized
lattice is a square lattice made of smaller squares. The intra-plaquette
interactions in such lattices might be strong and may even be designed to
include many($>2$)-body terms, while the inter-plaquette interactions might be
much weaker. Polymerized lattices allow for instance to engineer Valence Bond
Solids on demand, to study topological spin liquids and one might envisage
them as potential quantum circuits. It is also very appealing to try to use
the plaquettes as qubits or qudits (elementary systems with more than two
internal states) for quantum information processing and implement quantum
logical gates, quantum protocols, and quantum error correction in such systems
by employing either interatomic interactions or/and interactions with external
(electric, magnetic, laser) fields.

Here we take advantage of the 2 dimensional superlattices to present new
schemes to prepare universal 2D cluster states using the plaquettes as logical
qubits. The superlattices creates a periodic array of \emph{plaquettes}, i.e.,
$2\times2$ potential wells (as shown in Fig.~\ref{fig:Superlattice}), each of
them filled with an atom with two internal degrees of freedom (spin 1/2
particle). On each plaquette, we encode a single logical qubit using the
\emph{two-fold singlet subspace} of the four 1/2-spins, as shown in
Fig.~\ref{fig:Superlattice}. Thus, by doubling the physical resources in
comparison with the 2 physical qubit encoding previously mentioned, we obtain
the desired encoding that is decoherence free against uniform magnetic field
fluctuations in arbitrary directions.

The encoding scheme using the singlet subspace of four qubits has been
previously studied for the quantum dot systems, and it is also called the
\textquotedblleft supercoherent qubit\textquotedblright%
\ \cite{Bacon01,Weinstein05,Weinstein07}. Also, it has been shown that in such
a configuration, tunable Heisenberg superexchange interactions (between
neighboring spins, including the diagonal and off diagonal ones) are
sufficient for universal quantum computation \cite{Bacon00,Divincenzo00}. In
the model we propose here, the generation of a universal cluster state
demands: (i) the ability to perform one qubit gates to prepare all logical
qubits in the initial state $|+\rangle=1/\sqrt{(}2)[|0\rangle+|1\rangle]$, and
(ii) the realization of controlled-phase gates, $U=\mathrm{diag}(1,1,1,-1)$,
between nearest logical qubits, i.e. between plaquettes, to create a maximally
entangled 2D cluster state. Notice that the superlattice geometry of
Fig.~\ref{fig:Superlattice} does not induce interactions along the diagonals
sites on each plaquette, and thus, the most challenging ingredient for
universality is, indeed, the two qubit (two plaquette) gate. In this article,
we propose three different approaches to couple the logical qubits that either
preserve the singlet subspace at the end of the gate operation or keep the
state within the singlet subspace even during the whole completion of the
gate. In all approaches, we take into account realistic available tools and
discuss the practical limitations in optical superlattices. In our first
approach, we exploit the additional vibrational mode of the optical trap to
facilitate the logical coupling gate. In our second approach, we extend the
earlier proposals \cite{Weinstein05,Weinstein07} by removing the requirement
of equal coupling strengths for all six pairs within the plaquette (more
feasible for 2D optical superlattices), while we still obtain the effective
Hamiltonian within the logical subspace sufficient for universal gates. In our
last approach, we include tunable Ising-type interactions between neighboring
spins (attainable with neutral atoms in optical lattices \cite{Duan03}) and
use the optimal control techniques to find efficient and robust pulse
sequences for the logical coupling gate.

We notice that other proposals which exploit the superlattice structure in 2D
to create a universal resource (which is different from the universal cluster
state) for MBQC by connecting Bell entangled pairs by entangling phase gate
have been proposed recently \cite{Vaucher08}.



The paper is organized as follows: first, in Sec.~\ref{sec:ClusterState} we
present the general ideas for generating cluster states within a decoherence
free subspace (DFS) using optical superlattices. Then, in
Sec.~\ref{sec:Plaquette} we briefly review the singlet DFS of the plaquette
and describe operations of single logical qubit using superexchange couplings.
Finally, in Sec.~\ref{sec:Inter-Plaquette} we consider the key challenge of
implementing the logical controlled-phase gate with the singlet DFS encoding.
We propose three new approaches for the controlled-phase gate: the geometric
phase approach, the perturbative approach, and the optimal control approach. A
detailed comparison among the three approaches is summarized at the end of
this section (Table~\ref{tab:Comparison}) before we present our conclusion in
Sec.~\ref{sec:Conclusion}.

\section{Decoherence Free Cluster States \label{sec:ClusterState}}

One promising approach to quantum computation is the measurement-based quantum
computation (MBQC) \cite{Raussendorf01,Briegel01}, which uses universal
resources such as the cluster states. The cluster states can be efficiently
prepared by initializing all lattice spins in the product state of $\left\vert
+\right\rangle =\frac{1}{\sqrt{2}}\left(  \left\vert 0\right\rangle
+\left\vert 1\right\rangle \right)  $ and performing the
\emph{controlled-phase gate} between all pairs of neighboring spins. The
controlled-phase gate induces an additional factor $-1$ if both input qubits
are in state $\left\vert 1\right\rangle $. Up to some individual qubit
rotations, the controlled-phase gate can be achieved by Ising-type interaction
between two input qubits.

The preparation of a cluster state has been demonstrated using optical
lattices \cite{Mandel03}, with logical qubits directly stored in individual
spins. Controlled-collisions are used to implement the controlled-phase gate
between neighboring spins. However, such atomic qubits for the
controlled-collision scheme are vulnerable to magnetic field fluctuations,
which limits the practical implementation of the MBQC.


\begin{figure}[t]
\begin{center}
\includegraphics[
width=6.5 cm
]{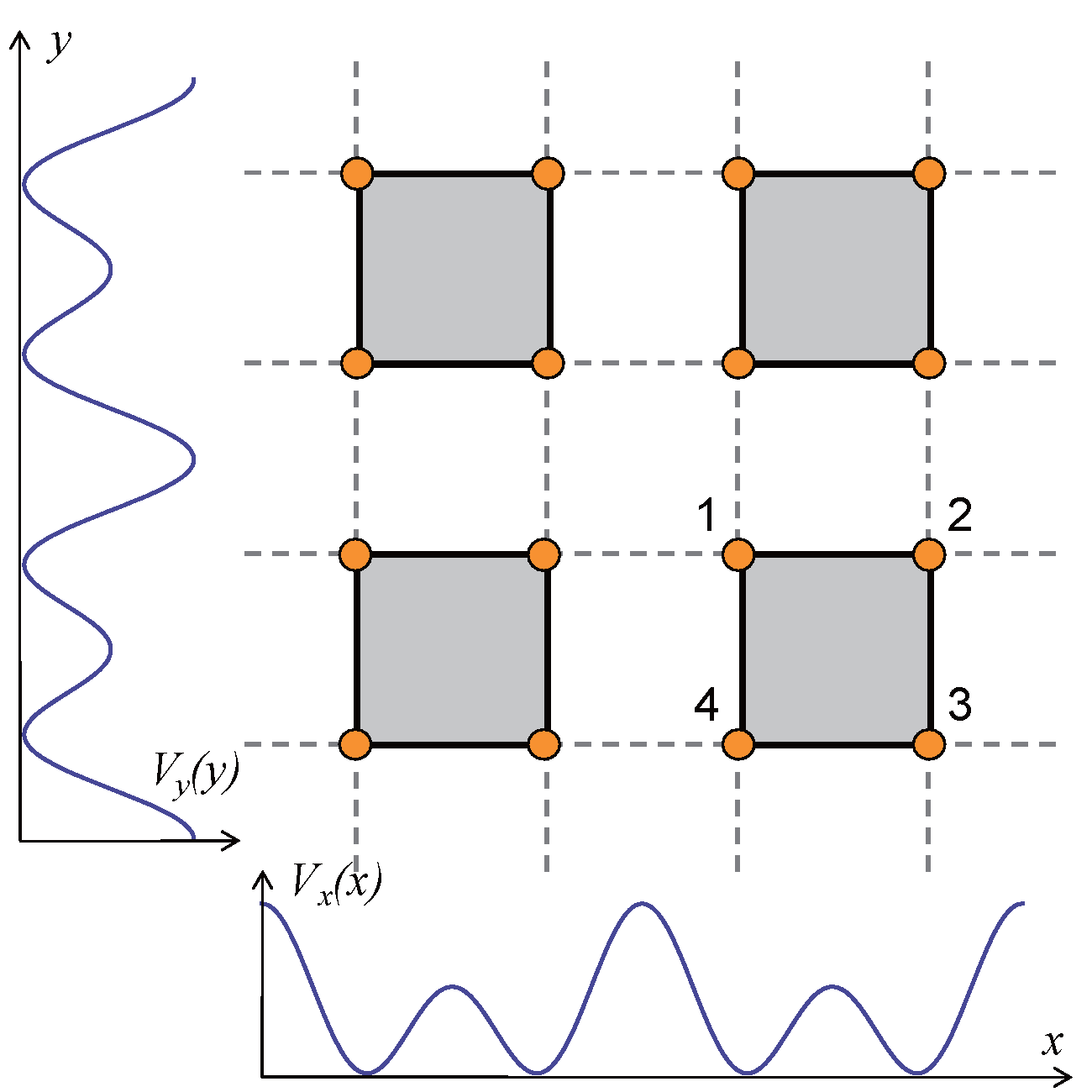}
\end{center}
\caption[fig:Superlattice]{(Color online) The optical superlattice consists of
a periodic array of $2\times2$ plaquettes. The 2D optical trapping potential
is created by adding two superlattice potentials $V_{x}\left(  x\right)  $ and
$V_{y}\left(  y\right)  $ illustrated in the bottom and left panels. The
intra-plaquette coupling is represented by the solid lines and the
inter-plaquette coupling is represented by the dashed lines.}%
\label{fig:Superlattice}%
\end{figure}

The optical superlattice inducing a periodic array of $2\times2$ plaquettes
(as shown in Fig.~\ref{fig:Superlattice}) can be created by superimposing two
optical lattice potentials with short and long wavelengths differing by a
factor of two \cite{Folling07} along both, x and y, directions. The effective
optical trapping potential becomes%
\begin{equation}
V=V_{x}\left(  x\right)  +V_{y}\left(  y\right)
\end{equation}
where $V_{u}=V_{u,s}\cos^{2}\left(  \frac{2\pi}{\lambda}u-\phi_{u,s}\right)
+V_{u,l}\cos^{2}\left(  \frac{\pi}{\lambda}u-\phi_{u,l}\right)  $ for $u=x,y$.
The short-lattice wavelength is $\lambda$, and the parameters $V_{u,s},$
$V_{u,l}$, $\phi_{u,s}$ and $\phi_{u,l}$ are controlled by the intensities and
phases of the laser beams.

For integer filling with one particle per site, each plaquette has four
particles. The four spin-1/2 particles have a two-fold singlet subspace with
total spin zero along all directions (i.e., $S_{tot}=0$). Thus, the singlet
subspace is the DFS insensitive to uniform magnetic field fluctuations.

The \emph{intra-plaquette couplings} (solid lines in
Fig.~\ref{fig:Superlattice}) enable operations of single logical qubit encoded
in the plaquette. (Note that by manipulating intra-plaquette couplings minimum
instances of topological matter can be demonstrated in the same optical
superlattice \cite{Paredes08}.) In order to create the decoherence free
cluster states, we will also need \emph{inter-plaquette couplings} (dashed
lines in Fig.~\ref{fig:Superlattice}) to implement the controlled-phase gates.

\begin{figure}[t]
\begin{center}
\includegraphics[
width=8.5 cm
]{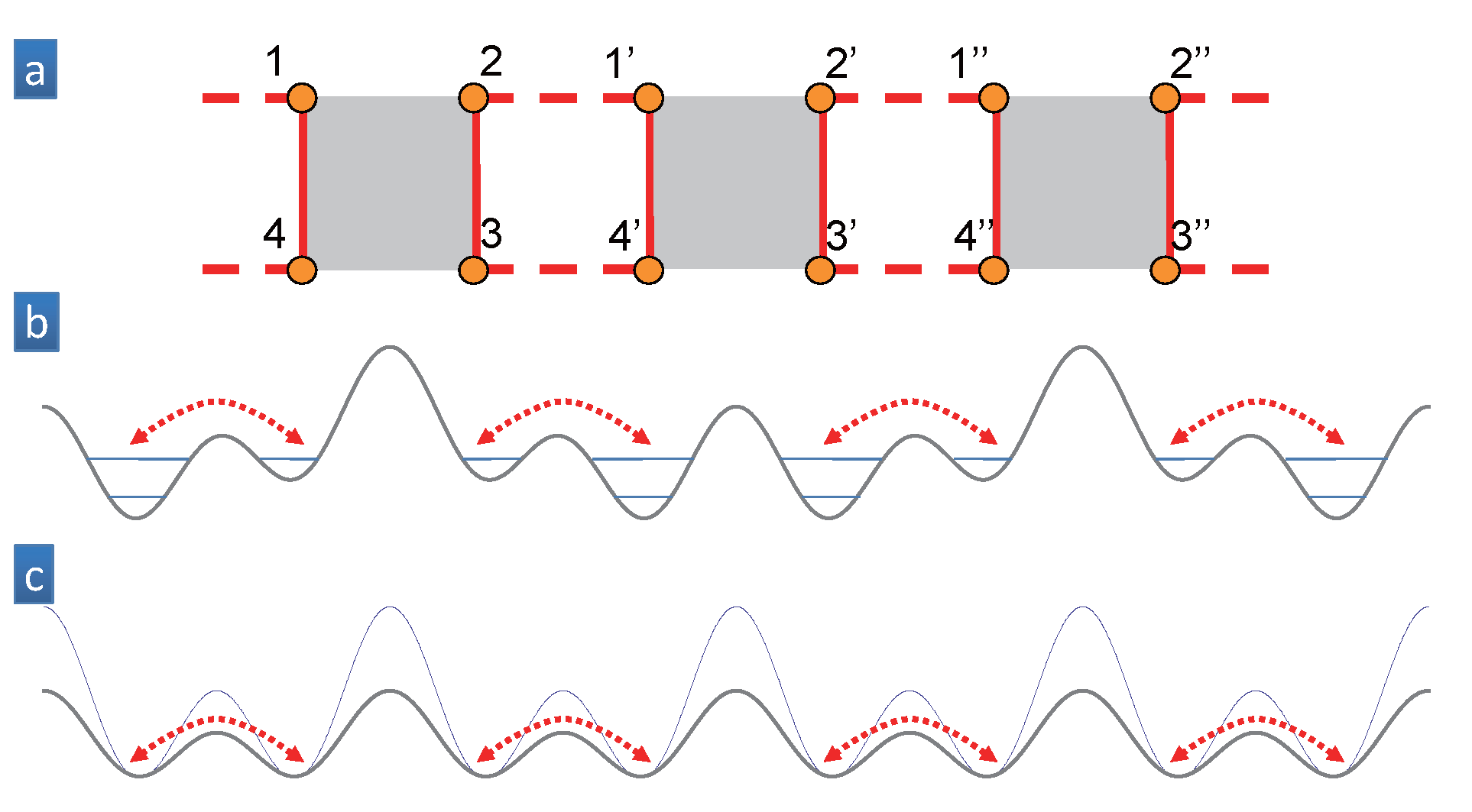}
\end{center}
\caption[fig:ClusterStateMulti]{(Color online) Simultaneous coupling between
neighboring plaquettes. (a) It is possible to simultaneously implement the
controlled-phase gates between neighboring plaquettes along the horizontal
direction, if only four sites are involved for each controlled-phase gate
(e.g., for the geometric phase approach or the optimal control approach). (b)
The optical superlattice (with inter-plaquette coupling and alternating energy
off-set for odd and even plaquettes, created by $\lambda,2\lambda,4\lambda$)
can yield the inter-plaquette coupling for the geometric phase approach. (c)
The spin-dependent optical superlattice (indicated by thin and thick lines of
potential profiles) can generate Ising interaction \cite{Duan03} useful for
the optimal control approach.}%
\label{fig:ClusterStateMulti}%
\end{figure}

There are eight sites for two neighboring plaquettes. If only four middle
sites are involved for the controlled-phase gate (e.g., for the geometric
phase approach in section~\ref{sec:GeoPhaseApproach} or the optimal control
approach in section~\ref{sec:OptimalControlApproach}), it is possible to
simultaneously apply controlled-phase gates to couple all horizontal (or
vertical) neighboring plaquettes with no overlap of sites involved for
different controlled-phase gates as shown in Fig.~\ref{fig:ClusterStateMulti}.
Meanwhile, if all eight sites from both plaquettes are involved for the
controlled-phase gate (e.g., the perturbative approach in
section~\ref{sec:PerturbativeApproach}), two steps are needed to couple the
plaquettes along the horizontal direction: first couple each even plaquette
with the neighboring odd plaquette on the left, and then couple each even
plaquette with the neighboring odd plaquette on the right.

In order to use the prepared cluster state for the MBQC, we should also be
able to measure the individual qubits. This can be achieved by first
converting the spin singlet/triplet states into different particle number
configurations \cite{Anderlini07,Folling07}, and then using various techniques
of coherent optical control with subwavelength resolution
\cite{Zhang06b,Cho07,Gorshkov08} to projectively count the particle number at
a specific site (without compromising the coherence for the remaining sites).

In the next two sections, we will consider the rotation of single logical
qubit using intra-plaquette couplings, and the controlled-phase gate between
two logical qubits using the additional inter-plaquette couplings, respectively.

\section{Logical Qubit Encoded in the Plaquette \label{sec:Plaquette}}

In this section, we focus on the operations within the plaquette via
\emph{intra-plaquette coupling}. For concreteness, we consider bosonic
particles, and similar results can be obtained for fermionic particles as
well. The following Hubbard Hamiltonian governs the dynamics of a single
plaquette, with spin-independent tunnelings and interactions (we will
introduce an additional vibration level in Sec.~IV):%
\begin{align}
H  &  =-\sum_{\left\langle i,j\right\rangle ,\sigma}\left(  t_{ij}a_{i\sigma
}^{\dag}a_{j\sigma}+h.c.\right)  +\frac{1}{2}U\sum_{i}n_{i}\left(
n_{i}-1\right) \nonumber\\
&  +\sum_{i,\sigma}\mu_{i}n_{i\sigma}%
\end{align}
where $a_{i\sigma}$ ($a_{i\sigma}^{\dag}$) is the annihilation (creation)
operator, $n_{i\sigma}$ is the particle number operator for site
$i=1,\cdots,4$ with spin $\sigma=\uparrow,\downarrow$, and $n_{i}%
=n_{i,\uparrow}+n_{i,\downarrow}$. The tunneling amplitudes ($t_{H}%
=t_{12}=t_{34}$ and $t_{V}=t_{23}=t_{41}$) and the offset energies ($\mu_{i}$)
can be changed by tuning the superlattice parameters. The large on-site
interaction $U\gg t_{H},t_{V}$ ensures that the system is in the Mott
insulator regime with fixed particle number for each site.

\begin{figure}[t]
\begin{center}
\includegraphics[
width=7 cm
]
{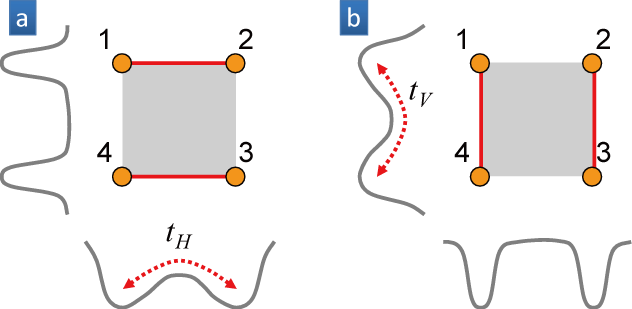}
\end{center}
\caption[fig:Plaquette]{(Color online) Intra-plaquette superexchange
couplings. The coupling strengths are (a) $J_{H}=t_{H}^{2}/U$ between
horizontal neighbors and (b) $J_{V}=t_{V}^{2}/U$ between vertical neighbors.
They can be changed independently by tuning the barriers between the sites.}%
\label{fig:Plaquette}%
\end{figure}

Particle tunneling only occurs virtually between neighboring sites, which
leads to the \emph{superexchange coupling}
\begin{equation}
H_{eff}=-J_{H}\left(  \vec{s}_{1}\cdot\vec{s}_{2}+\vec{s}_{3}\cdot\vec{s}%
_{4}\right)  -J_{V}\left(  \vec{s}_{2}\cdot\vec{s}_{3}+\vec{s}_{4}\cdot\vec
{s}_{1}\right)  ,
\end{equation}
where $\vec{s}_{i}$ are Pauli operators for the spin at site $i$. The coupling
strengths $J_{H}=t_{H}^{2}/U$ and $J_{V}=t_{V}^{2}/U$ can be changed
independently, by tuning the barriers between the sites as illustrated in
Fig.~\ref{fig:Plaquette}. Controlling the superexchange couplings is
sufficient to perform arbitrary rotations of the logical qubit encoded in the plaquette.


\subsection{Singlet subspace for four spins}

The space of 4 1/2-spin particles span a subspace of total spin 2, 3 subspaces
of total spin 1 and 2 subspaces of total spin 0. We use the two-fold
\emph{singlet subspace} of the plaquette to encode the logical qubit. The
singlet subspace is spanned by%
\begin{equation}
\left\vert \Psi_{H}\right\rangle =\left\vert S\right\rangle _{1,2}%
\otimes\left\vert S\right\rangle _{3,4}%
\end{equation}%
\begin{equation}
\left\vert \Psi_{V}\right\rangle =\left\vert S\right\rangle _{2,3}%
\otimes\left\vert S\right\rangle _{4,1},
\end{equation}
with $\left\vert S\right\rangle _{i,j}\equiv\frac{1}{\sqrt{2}}\left(
\left\vert \uparrow\right\rangle _{i}\left\vert \downarrow\right\rangle
_{j}-\left\vert \downarrow\right\rangle _{i}\left\vert \uparrow\right\rangle
_{j}\right)  $. $\left\vert \Psi_{H}\right\rangle $ (or $\left\vert \Psi
_{V}\right\rangle $) is the product state of two singlet pairs along the
horizontal (or vertical) direction, which can be prepared using the procedure
demonstrated in \cite{Trotzky08}.

The singlet subspace is decoherence free, because it is insensitive to the
uniform magnetic field fluctuations. In addition, measuring a single spin will
not distinguish the states from singlet subspace, and this is a source of
protection against local perturbations. Since $\left\langle \Psi_{H}|\Psi
_{V}\right\rangle =1/2\neq0$, it is more convenient to use the orthogonal
states $\left\vert 0\right\rangle \equiv\left\vert \Psi_{V}\right\rangle $ and
$\left\vert 1\right\rangle \equiv\frac{2}{^{\sqrt{3}}}\left(  \frac{1}%
{2}\left\vert \Psi_{V}\right\rangle -\left\vert \Psi_{H}\right\rangle \right)
$. We can also write the orthogonal states in terms of the singlets and
triplets for vertical pairs $\left(  2,3\right)  $ and $\left(  4,1\right)  $
\cite{Bacon00}:%

\begin{equation}
\left\vert 0\right\rangle =\left\vert S\right\rangle _{2,3}\left\vert
S\right\rangle _{4,1}%
\end{equation}%
\begin{equation}
\left\vert 1\right\rangle =\frac{1}{\sqrt{3}}\left[  \left\vert T_{+}%
\right\rangle _{2,3}\left\vert T_{-}\right\rangle _{4,1}-\left\vert
T_{0}\right\rangle _{2,3}\left\vert T_{0}\right\rangle _{4,1}+\left\vert
T_{-}\right\rangle _{2,3}\left\vert T_{+}\right\rangle _{4,1}\right]
\end{equation}
where $\left\vert T_{+,0,-}\right\rangle =\left\{  \left\vert \uparrow
\right\rangle \left\vert \uparrow\right\rangle ,\frac{1}{\sqrt{2}}\left(
\left\vert \uparrow\right\rangle \left\vert \downarrow\right\rangle
+\left\vert \downarrow\right\rangle \left\vert \uparrow\right\rangle \right)
,\left\vert \downarrow\right\rangle \left\vert \downarrow\right\rangle
\right\}  $. For such a choice of basis, the subsystem of two spins $\left(
2,3\right)  $ is sufficient to determine the logical states $\left\vert
0\right\rangle $ and $\left\vert 1\right\rangle $, because the corresponding
reduced density matrices%

\begin{align}
\rho_{2,3}^{\left\vert 0\right\rangle }  &  =\operatorname*{Tr}\nolimits_{4,1}%
\left[  \left\vert 0\right\rangle \left\langle 0\right\vert \right]  =\left(
\left\vert S\right\rangle \left\langle S\right\vert \right)  _{2,3}\\
\rho_{2,3}^{\left\vert 1\right\rangle }  &  =\operatorname*{Tr}\nolimits_{4,1}%
\left[  \left\vert 1\right\rangle \left\langle 1\right\vert \right]  =\frac
{1}{3}\left(  \left\vert T_{+}\right\rangle \left\langle T_{+}\right\vert
+\left\vert T_{0}\right\rangle \left\langle T_{0}\right\vert +\left\vert
T_{-}\right\rangle \left\langle T_{-}\right\vert \right)  _{2,3},
\end{align}
belong to orthogonal (singlet and triplet) subspaces, $\operatorname*{Tr}%
\left[  \rho_{2,3}^{\left\vert 0\right\rangle }\rho_{2,3}^{\left\vert
1\right\rangle }\right]  =0$.

The Pauli operators associated with the logical qubit are: $\sigma_{x}%
\equiv\left\vert 0\right\rangle \left\langle 1\right\vert +\left\vert
1\right\rangle \left\langle 0\right\vert $, $\sigma_{y}\equiv i\left\vert
0\right\rangle \left\langle 1\right\vert -i\left\vert 1\right\rangle
\left\langle 0\right\vert $, and $\sigma_{z}\equiv\left\vert 0\right\rangle
\left\langle 0\right\vert -\left\vert 1\right\rangle \left\langle 1\right\vert
$. Within the singlet subspace the logical operator $\sigma_{z}$ can be
achieved by operating the $\left(  2,3\right)  $ spins%
\begin{equation}
\sigma_{z}\doteq-\frac{1}{2}\left(  1+\vec{s}_{2}\cdot\vec{s}_{3}\right)  ,
\label{eq:SigmaZ23}%
\end{equation}
where we use "$\doteq$" to represent the special equality valid within the
singlet subspace. Since $\vec{s}_{2}\cdot\vec{s}_{3}\doteq\vec{s}_{4}\cdot
\vec{s}_{1}$ (i.e., either both pairs are singlets or both are triplets),
$\sigma_{z}$ can also be implemented by operating the $\left(  4,1\right)  $
spins%
\begin{equation}
\sigma_{z}\doteq-\frac{1}{2}\left(  1+\vec{s}_{4}\cdot\vec{s}_{1}\right)  .
\label{eq:SigmaZ41}%
\end{equation}

\subsection{Rotating logical qubit with super-exchange coupling}

\begin{figure}[t]
\begin{center}
\includegraphics[
width=8 cm
]
{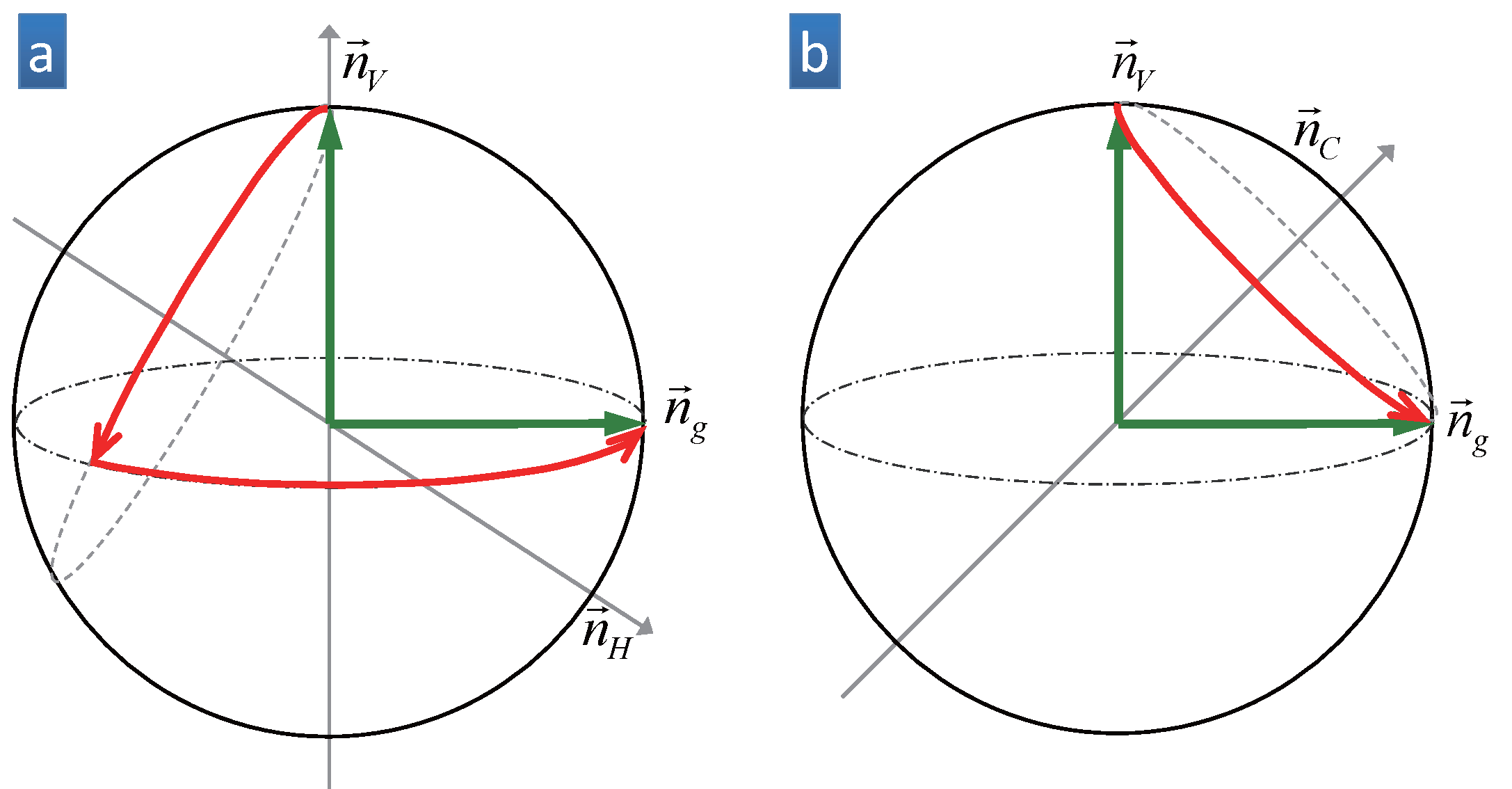}
\end{center}
\caption[fig:BlochSphere]{(Color online) The Bloch sphere representation for
the singlet subspace. The state $\left\vert 0\right\rangle $ ($\left\vert
1\right\rangle $) is associated with the north (south) pole. (a) The
superexchange coupling $J_{H}$ ($J_{V}$) is associated with the rotation
around the $\vec{n}_{H}$ ($\vec{n}_{V}$) axis. Here $\vec{n}_{H}=\left(
\frac{\sqrt{3}}{2},0,\frac{-1}{2}\right)  $ and $\vec{n}_{V}=\left(
0,0,1\right)  $. The sequential rotations around the $\vec{n}_{H}$ and
$\vec{n}_{V}$ axes can rotate the Bloch vector from $\vec{n}_{V}$ to $\vec
{n}_{g}=\left(  \frac{1}{\sqrt{2}},0,\frac{1}{\sqrt{2}}\right)  $ [i.e., state
$\frac{1}{\sqrt{2}}\left(  \left\vert 0\right\rangle +\left\vert
1\right\rangle \right)  $]. (b) Alternatively combined superexchange coupling
(with contributions from both $J_{H}$ and $J_{V}$) can implement the rotation
around the axis $\vec{n}_{C}$, which rotates the Bloch vector from $\vec
{n}_{V}$ to $\vec{n}_{g}$ in one step.}%
\label{fig:BlochSphere}%
\end{figure}

We now consider arbitrary rotations in the singlet subspace using
super-exchange couplings. First of all, the superexchange coupling Hamiltonian
commutes with the total spin operator of the plaquette $\left[  H_{eff}%
,\vec{s}_{1}+\vec{s}_{2}+\vec{s}_{3}+\vec{s}_{4}\right]  =0$, due to the
identity $\left[  \vec{s}_{i}\cdot\vec{s}_{j},\vec{s}_{i}+\vec{s}_{j}\right]
=0$. Consequently, the superexchange coupling preserves the singlet subspace
(with zero total spin). Within the singlet subspace we have%
\begin{align}
\vec{n}_{H}\cdot\vec{\sigma} &  \doteq\frac{1}{2}-\frac{1}{4}\left(  \vec
{s}_{1}\cdot\vec{s}_{2}+\vec{s}_{3}\cdot\vec{s}_{4}\right)  \\
\vec{n}_{V}\cdot\vec{\sigma} &  \doteq\frac{1}{2}-\frac{1}{4}\left(  \vec
{s}_{2}\cdot\vec{s}_{3}+\vec{s}_{4}\cdot\vec{s}_{1}\right)
\end{align}
with $\vec{n}_{H}=\left(  \frac{\sqrt{3}}{2},0,\frac{-1}{2}\right)  $ and
$\vec{n}_{V}=\left(  0,0,1\right)  $ as illustrated in
Fig.~\ref{fig:BlochSphere}(a). The constant of $1/2$ can be neglected, as it
only induces an overall phase during the evolution.
The rotations about these axes can be controlled by switching on/off the
superexchange couplings of $J_{H}$ and $J_{V}$, which varies expentially with
the height of the corresponding barriers.
Since the angle between $\vec{n}_{H}$ and $\vec{n}_{V}$ is $2\pi/3$, arbitrary
rotation of the Bloch sphere can be achieved within $4$ operations. (This is a
special case of the general theorem \cite{Lowenthal71, Khaneja07} stating that
$k+2$ operations are sufficient for arbitrary rotation given the angle $\eta$
between the two rotation axes satisfies $\frac{\pi}{k}>\min\left(  \eta
,\pi-\eta\right)  \geq\frac{\pi}{k+1}$.)

The product state of two vertical singlet pairs $\left\vert 0\right\rangle
=\left\vert \Psi_{V}\right\rangle $ can be initialized using the procedure
demonstrated in \cite{Trotzky08}. Universal rotation enables dynamical
preparation of arbitrary logical state encoded in singlet subspace. For
example, $\left\vert +\right\rangle =\frac{1}{\sqrt{2}}\left(  \left\vert
0\right\rangle +\left\vert 1\right\rangle \right)  $ can be prepared by
two-step evolution $e^{-i\vec{n}_{V}\cdot\vec{\sigma}\theta_{V}}e^{-i\vec
{n}_{H}\cdot\vec{\sigma}\theta_{H}}\left\vert \Psi_{V}\right\rangle $ with
$2\theta_{H}=2\sin^{-1}\frac{\sqrt{2}}{\sqrt{3}}\approx109.5^{\circ}$ and
$2\theta_{V}=\pi-\sin^{-1}\frac{\sqrt{2}}{\sqrt{3}}\approx125.3^{\circ}$, as
shown in Fig.~\ref{fig:BlochSphere}(a). Alternatively, we can tune the
relative strength between $\vec{n}_{H}\cdot\vec{\sigma}$ and $\vec{n}_{V}%
\cdot\vec{\sigma}$ to achieve the total coupling $\vec{n}_{C}\cdot\vec{\sigma
}$ with $\vec{n}_{C}=\left(  \frac{1}{\sqrt{2}},0,\frac{1}{\sqrt{2}}\right)
$, and prepare $\left\vert +\right\rangle $ in one step $e^{-i\vec{n}_{C}%
\cdot\vec{\sigma}\frac{\pi}{2}}\left\vert \Psi_{V}\right\rangle $, as
illustrated in Fig.~\ref{fig:BlochSphere}(b). Note that all the plaquettes can
be simultaneously prepared in the $\left\vert +\right\rangle $ state. In order
to create the decoherence free cluster state, we need the controlled-phase
gate between the logical qubits encoded in neighboring plaquettes.

\section{Controlled-Phase Gate \label{sec:Inter-Plaquette}}

We now consider inter-plaquette couplings (dashed lines in
Fig.~\ref{fig:Superlattice}). In particular, we focus on implementing the
controlled-phase gate between two neighboring plaquettes, which induces an
additional $-1$ phase if both encoded qubits are in the logical state
$\left\vert 1\right\rangle $. In principle, the controlled-phase gate can be
achieved by the Ising-type interaction between the logical qubits, but
unfortunately such interaction is not immediately available from the lattice
experiments, as the effective Ising term $\sigma_{z}\sigma_{z}^{\prime}$
requires four-site interaction $\left(  \vec{s}_{2}\cdot\vec{s}_{3}\right)
\left(  \vec{s}_{4}^{\prime}\cdot\vec{s}_{1}^{\prime}\right)  $ [see
Eqs.~(\ref{eq:SigmaZ23},\ref{eq:SigmaZ41})]. However, since what we want is
the specific unitary evolution rather than the interaction, it is actually
more feasible to implement the unitary evolution directly.

In the following, we present three different approaches to implement the
controlled-phase gate between two neighboring plaquettes. For concreteness, we
only consider coupling two neighboring plaquettes along the horizontal
direction, while all three approaches can also couple neighboring plaquettes
along the vertical direction.

\subsection{Geometric phase approach\label{sec:GeoPhaseApproach}}

\begin{figure}[t]
\begin{center}
\includegraphics[
width=7.6 cm
]
{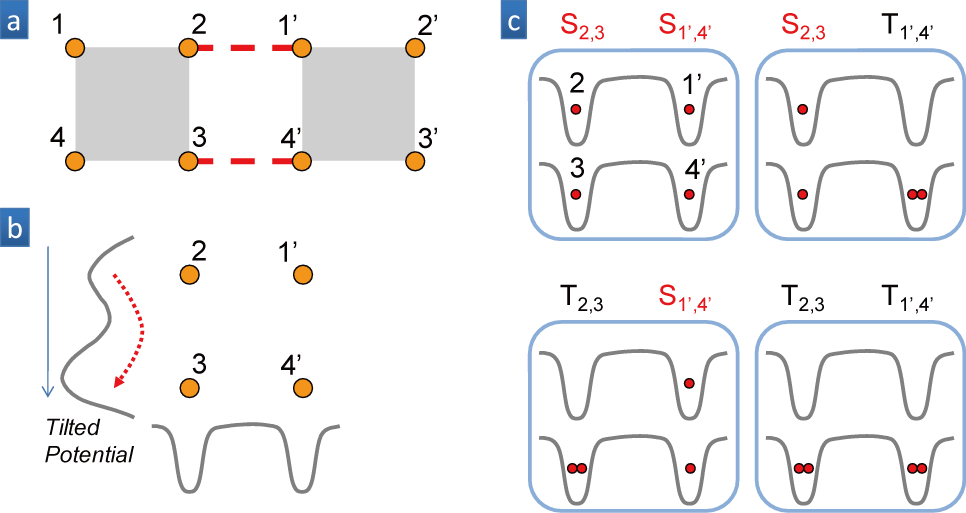}
\end{center}
\caption[fig:GeoPhaseGate4Boson1]{(Color online) Geometric phase approach to
controlled-phase gate (\textbf{Step~1}). (a) The sites from two neighboring
plaquettes are labeled. (b) The intra-plaquette trapping potential along the
vertical direction is adiabatically tilted. This results in single (or double)
occupancy at the lower site if the vertical pair of particles is in the
singlet (or triplet) state. (c) Particle number configurations are plotted for
four possible of spin states: $S_{2,3}\otimes S_{1^{\prime},4^{\prime}}$,
$S_{2,3}\otimes T_{1^{\prime},4^{\prime}}$, $T_{2,3}\otimes S_{1^{\prime
},4^{\prime}}$, and $T_{2,3}\otimes T_{1^{\prime},4^{\prime}}$. ($S$ and $T$
indicate singlet and triplet.)}%
\label{fig:GeoPhaseGate4Boson1}%
\end{figure}

The first approach uses the vibration levels and the geometric phase to
achieve the controlled-phase gate between neighboring plaquettes
[Fig.~\ref{fig:GeoPhaseGate4Boson1}(a)]. The geometric phase is proportional
the surface area enclosed by the evolution trajectory in the Bloch sphere
(associated with the two energy levels that are degenerate). For example, if a
half of the Bloch sphere is enclosed, the system acquires a geometric phase
$\pi$. We first consider the bosonic particles. It takes three steps to
achieve the controlled-phase gate:

\textbf{Step 1.} We lower the inter-plaquette barrier and adiabatically tilt
the intra-plaquette potential along the vertical direction
[Fig.~\ref{fig:GeoPhaseGate4Boson1}(b)]. Each lower site will be occupied by
one particle (or two particles) if the vertical pair of particles is in the
singlet (or triplet) state [Fig.~\ref{fig:GeoPhaseGate4Boson1}(c)]. For
example, if the spins $\left(  2,3\right)  $ are in the singlet state (denoted
as $S_{2,3}$), the transfer of particle from site $2$ to site $3$ is prevented
by the symmetry requirement of bosonic particles, resulting in one particle in
site $3$ [see the upper two panels in Fig.~\ref{fig:GeoPhaseGate4Boson1}(c)].
If the spins $\left(  2,3\right)  $ are in the triplet subspace (denoted as
$T_{2,3}$), the particle from site $2$ is adiabatically transferred to site
$3$, leaving two particles in site $3$ [see the lower two panels in
Fig.~\ref{fig:GeoPhaseGate4Boson1}(c)]. Similar spin-dependent transfer also
happens to other sites, such as $\left(  1^{\prime},4^{\prime}\right)  $.

\begin{figure}[t]
\begin{center}
\includegraphics[
width=8.5 cm
]
{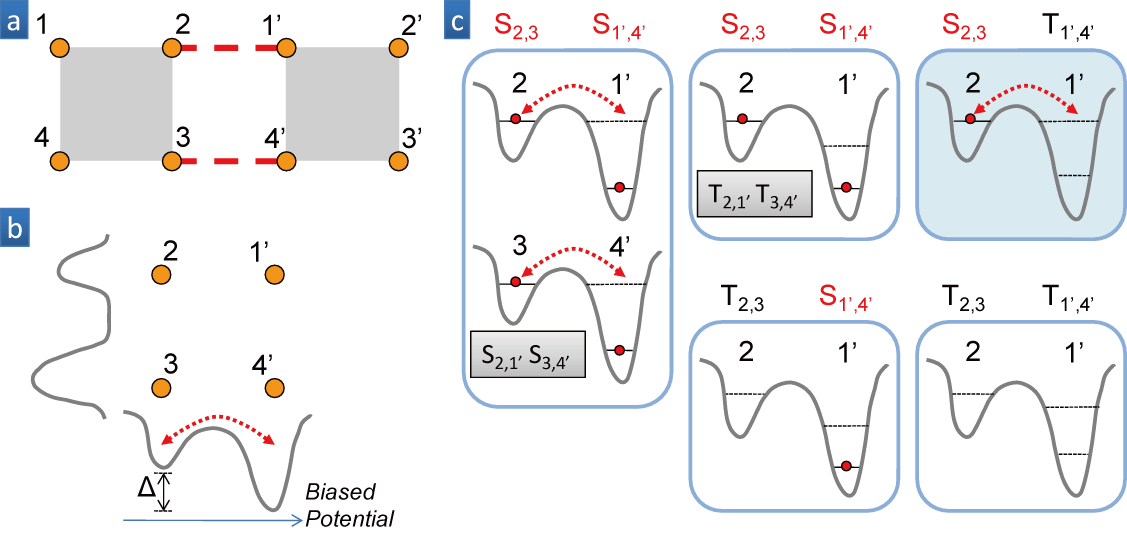}
\end{center}
\caption[fig:GeoPhaseGate4Boson1]{(Color online) Geometric phase approach to
controlled-phase gate (\textbf{Step~2}). (a) The sites from two neighboring
plaquettes are labeled. (b) A defined bias ($\Delta$) of the inter-plaquette
potential is quickly applied and the inter-plaquette barrier is lowered to
facilitate the resonant tunneling along the horizontal direction. (c) Resonant
tunneling (between the vibrational ground level of the left site and the
vibrational excited level of the right site) can occur for the following two
cases: 1) each of the left and right sites has exactly one particle, and the
two particles are in the singlet state (see the left panel), 2) the left site
has one particle and the right site has zero particle (see the highlighted
upper right panel). All other configurations are off-resonant, with negligible
tunneling. After time $2\pi/t$, a geometric phase $\pi$ from the resonant
tunneling is obtained for $S_{2,3}\otimes T_{1^{\prime},4^{\prime}}$ (the
highlighted upper right panel), while only a trivial geometric phase ($0$ or
$2\pi$) is obtain for the other three cases ($S_{2,3}\otimes S_{1^{\prime
},4^{\prime}}$, $T_{2,3}\otimes S_{1^{\prime},4^{\prime}}$, and $T_{2,3}%
\otimes T_{1^{\prime},4^{\prime}}$).}%
\label{fig:GeoPhaseGate4Boson2}%
\end{figure}

\textbf{Step 2. }We quickly apply a defined bias to the inter-plaquette
lattice potential and lower the inter-plaquette barrier along the horizontal
direction [Fig.~\ref{fig:GeoPhaseGate4Boson2}(b)]. This induces single
particle resonant tunneling with rate $t$ between the vibrational ground state
at site $2$ and the vibrational excited state at site $1^{\prime}$
\footnote{In general, the tunneling rate between two sites in different
vibrational bands is zero for an untilted lattice. However, here we have a
finite rate $t$ because the Wannier orbitals are distorted for the tilted
lattice.}, if there is one particle at site $2$ and zero particle at site
$1^{\prime}$ [see the highlighted upper right panel in
Fig.~\ref{fig:GeoPhaseGate4Boson2}(c)]. By waiting for time $2\pi/t$, we
obtain the geometric phase $\pi$ from the resonant tunneling for
$S_{2,3}\otimes T_{1^{\prime},4^{\prime}}$. As detailed in Sec.~IV, for all
other three cases ($S_{2,3}\otimes S_{1^{\prime},4^{\prime}}$, $T_{2,3}\otimes
S_{1^{\prime},4^{\prime}}$, and $T_{2,3}\otimes T_{1^{\prime},4^{\prime}}$) we
only obtain a trivial geometric phase $0$ (or $2\pi$).

\textbf{Step 3. }We change the intra-plaquette potential to the initial
balanced position along the vertical direction (having one particle per site)
and restore each plaquette to the logical subspace.

A recent superlattice experiment uses the resonant tunneling and the blockade
induced by on-site interaction to count the number of atoms \cite{Cheinet08}.
This experiment demonstrates that the presence or absence of resonant
tunneling can be highly sensitive to the number of particles in the lattice
sites. The geometric phase approach can be regarded as an extension that uses
the resonant tunneling to coherently imprint a geometric phase for a specific
particle number configuration (corresponding to certain logical state).

The procedure for the fermionic particles is almost the same as that for the
bosonic particles, except for the following three differences. First, the bias
of the energy off-set needs to be $\Delta=\omega+U_{R}^{ab}$ for fermionic
particles (whereas $\Delta=\omega$ for bosonic particles), where $\omega$ is
the vibrational excitation energy and $U_{R}^{ab}$ is the on-site interaction
between ground and excited levels at the right site ($1^{\prime}$ or
$4^{\prime}$). Second, the geometric phase $\pi$ is obtained from the resonant
tunneling associated the subspace $T_{2,3}\otimes S_{1^{\prime},4^{\prime}}$
for fermionic particles (whereas it is associated with $S_{2,3}\otimes
T_{1^{\prime},4^{\prime}}$ for bosonic particles). Third, the geometric phase
is $0$ for the remaining cases for fermionic particles (whereas it might be
either $0$ or $2\pi$ for bosonic particles).

It is tempting to consider using $\Delta=\omega$ for the fermionic particles,
as we might expect that by exchanging the roles of singlet and triplets, the
fermionic particles could be mapped to bosonic particles. However, the roles
of singlet and triplets are not exactly symmetric. For example, consider the
case with one particle per site after Step 1. For bosonic particles, the
system is in the subspace $S_{2,3}\otimes S_{1^{\prime},4^{\prime}}$ that has
finite projection to $S_{2,1^{\prime}}\otimes S_{3,4^{\prime}}$ and
$T_{2,1^{\prime}}\otimes T_{3,4^{\prime}}$ (but not $T_{2,1^{\prime}}\otimes
S_{3,4^{\prime}}$ or $S_{2,1^{\prime}}\otimes T_{3,4^{\prime}}$), which yields
a trivial $0$ or $2\pi$ geometric phase. For fermionic particles, the system
is in the subspace $T_{2,3}\otimes T_{1^{\prime},4^{\prime}}$ that has finite
projection to $T_{2,1^{\prime}}\otimes S_{3,4^{\prime}}$ and $S_{2,1^{\prime}%
}\otimes T_{3,4^{\prime}}$, as well as $S_{2,1^{\prime}}\otimes S_{3,4^{\prime
}}$ and $T_{2,1^{\prime}}\otimes T_{3,4^{\prime}}$, which thus may yields a
non-trivial $\pi$ geometric phase.

The detailed calculation for both bosonic and fermionic particles are
presented in Appendix~\ref{app:GeoPhaseApproach}.

\subsection{Perturbative approach\label{sec:PerturbativeApproach}}


The second approach uses both the intra- and inter-plaquette couplings acting
on the eight sites. The intra-plaquette coupling induces an energy gap between
the logical states (i.e., singlet subspace)\ and other non-logical states,
while the inter-plaquette coupling acts as a perturbation that induces
different phase shifts for different logical states. The inter-plaquette
coupling can be efficiently achieved using super-exchange interaction between
the inter-plaquette neighboring sites.

The key challenge is to obtain the intra-plaquette interaction, with finite
Heisenberg interaction between the sites along the diagonal and off-diagonal
directions. We can overcome the challenge by using a different design of the
optical lattice.

\subsubsection{Lattice Geometry and Energy levels}

\begin{figure}[ptbh]
\centering
\centering
{\includegraphics[width=8.7 cm]{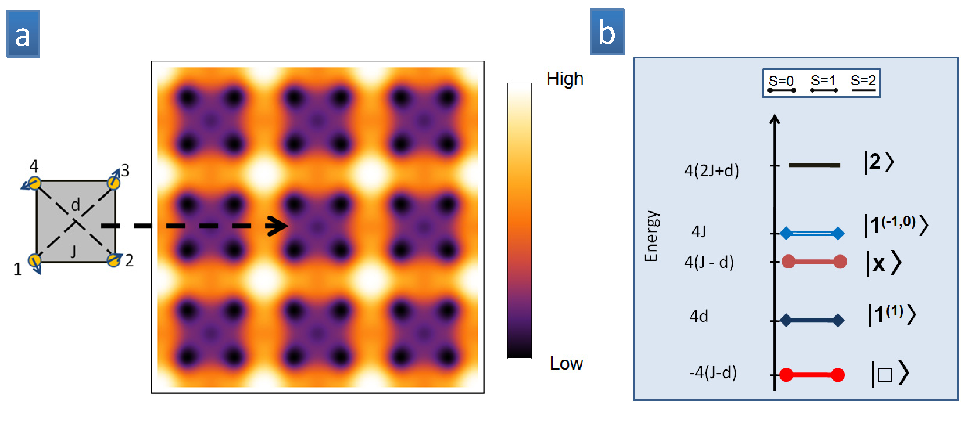}}
\caption[fig:PertLatticeLevels]{(Color online) (a) Density plot of the lattice
potential [see Eq.~(\ref{eq:lattice2})], which generates an array of
$2\times2$ plaquettes in the $x-y$ plane with periodicity of $2\pi/k$. Using
parameters $V^{l}=60E_{R}$, $V^{s}=9E_{r}$ and $V^{\prime}=10E_{r}$, with
$E_{R}$ and $Er$ photon recoil energy of the long and short lattices
respectively, one can achieve a parameter regime with $d/J\approx0.2$. (b)
Energy levels of a single plaquette described by the Hamiltonian given in
Eq.~(\ref{eq:effc}). In the plot we assume fermionic atoms,i.e. $J,d>0$
(fermions). For bosons $J,d<0$ the order of the energy levels is reverse i.e
the $S=2$ is the lowest energy state. }%
\label{fig:PertLatticeLevels}%
\end{figure}

We want to obtain the Hamiltonian with intra-plaquette interaction:%

\begin{equation}
H_{intra}=J\sum_{i=1,2,3,4}\vec{s}_{i}\cdot\vec{s}_{i+1}+d\sum_{i=1,2}\vec
{s}_{i}\cdot\vec{s}_{i+2}, \label{eq:effc}%
\end{equation}
where the exchange interaction $J=\pm t^{2}/U$ ($d=\pm\tilde{t}^{2}/U$) is
induced by the tunneling between the nearest neighbors (next-nearest
neighbors) with tunneling rate $t$ ($\tilde{t}$). The positive and negative
signs are for fermions and bosons, respectively. To simplify the notation, we
have identified $\vec{s}_{5}$ with $\vec{s}_{1}$. Such type of interaction can
be created by a lattice potential of the form [see
Fig.~\ref{fig:PertLatticeLevels}(a)]

\begin{equation}
V(x,y,z)=V^{c}(x,y)+V_{x}^{s}(x)+V_{y}^{s}(y),\label{eq:lattice2}%
\end{equation}
where $V_{u}^{s}(u)=\frac{V^{s}}{2}\cos(2ku)-\frac{V^{l}}{2}\cos(ku)$ are the
typical double well superlattice formed by the superposition of two
independent sinusoidal potentials which differ in periodicity, $a=\pi
/k,2\pi/k$, by a factor of two \cite{Trotzky08}
and $V^{c}(x,y)=-V^{\prime2}(k(x-y))\cos^{2}(k(x+y))$ is an additional
potential that allows to control the diagonal couplings within the plaquettes.
It can be constructed for example from a folded, retro-reflected beam with
out-of-the-plane polarization \cite{Sebby06}.
By varying the depths of the short, $V^{s}$ and $V^{l}$ long lattices it is
possible to control the intra and inter-plaquette coupling independently, and
in particular to make the latter negligibly small and the plaquettes independent.

As the intensity of the non-separable part of the potential is ramped up, a
minima at the center of the plaquettes develops. If the strength of the latter
is such that the energies of bound states in this minima are larger than the
energies of the lowest vibrational states at the plaquette sites it is
possible to tune the ratio $\tilde{t}/t$ without populating the central site,
which is required for the validity of Eq.~(\ref{eq:effc}). For example using
the parameters $V^{l}=60E_{R}=15E_{r}$, $V^{s}=9E_{r}$ and $V^{\prime}%
=10E_{r}$, with $E_{R}=\hbar^{2}k^{2}/(8m)$ and $Er=\hbar^{2}k^{2}/(2m)$ the
photon recoil energy of the long and short lattices respectively one can
achieve a parameter regime with $\tilde{t}/t\approx0.5$ with an energy gap to
the first vibrational state in the central well of order $E_{g}/t\approx10$.
It is very difficult to increase $\tilde{t}/t$ close to one by just
controlling the lattice potential, because the energy gap disappears and the
central sites become accessible. Therefore, we will focus on the case $d<J$.

The eigenstates
associated with Eq.~(\ref{eq:effc}) can be classified according with their
total spin $S$. As shown in Fig.~\ref{fig:PertLatticeLevels}(b), there are two
singlet ($S=0$) states
\begin{align}
\left\vert \Box\right\rangle  &  =\frac{1}{\sqrt{3}}(|\Psi_{H}\rangle
+|\Psi_{V}\rangle)\\
\left\vert \times\right\rangle  &  =|\Psi_{H}\rangle-|\Psi_{V}\rangle,
\end{align}
with energies $E(\left\vert \Box\right\rangle )=-4(J-d)$ and $E(\left\vert
\times\right\rangle )=+4(J-d)$, respectively. There are three $S=1$ states
denoted by $|1^{(-1,0,1)}\rangle$, with energies $E(1^{(q)})=4J,4J$ and $4d$,
respectively. For fermionic (bosonic) atoms the highest (lowest) energy state
is a $S=2$ state, $|2\rangle$ with energy $E(2)=4(2J+d)$.

We want to use the singlet states within each plaquette as encoded qubits and
perform a phase gate between them by coupling nearest neighbor plaquettes into
a \textquotedblleft\emph{superplaquette}\textquotedblright\ (i.e., $2\times4$
potential wells). A superplaquette can be achieve by superimposing laser beams
with periodicities $4\lambda$ and $4\lambda/3$ along one axis \cite{Rey08}.
Such wavelength are experimentally available for typical Alkali atoms or can
be engineered by intersecting pairs of laser beams at appropriated angles
\cite{Peil03}.
The $4\lambda$ isolates pairs of adjacent plaquettes along one direction and
the extra $4\lambda/3$ lattice is needed to balance the offset created when
the latter lattice is added. When pairs of plaquettes are weakly coupled into
a superplaquette the Hamiltonian that connects the plaquettes is given by
\begin{equation}
H_{c}=J^{\prime}(\vec{s}_{2}\cdot\vec{s}_{1^{\prime}}+\vec{s}_{3}\cdot\vec
{s}_{4^{\prime}}).
\end{equation}
We want to use the coupling to implement a controlled-phase gate between the
singlet eigenstates in the two plaquettes. To achieve that we require that the
inter-plaquette coupling is weak (i.e., $J^{\prime}\ll\min\{4d,8\left(
J-d\right)  ,4(J-2d)\}$) and derive an effective Hamiltonian by adiabatically
eliminating the all $S>0$ states.

In the following we discuss the implementation of the controlled-phase gate
for the experimentally relevant regime $d<J$. (The ideal case of $d=J$ is
discussed in Appendix~\ref{app:d&J}.)

\subsubsection{Perturbative approach with $d<J$}

For the 2D plaquette implementation the diagonal coupling $d$ is always
smaller than $J$ and therefore the singlet states within the plaquette are
non-degenerate
\begin{equation}
|\Delta E|=|E(\left\vert \times\right\rangle )-E(\left\vert \Box\right\rangle
)|=8|J-d|>0.
\end{equation}
Regardless of this issue, it is still possible to derive an effective
Hamiltonian provided that the inter-plaquette coupling $J^{\prime}$ is less
than the energy difference between $|\times\rangle$ and all other states [see
Fig.~\ref{fig:PertLatticeLevels}(b)]:
\begin{equation}
J^{\prime}\ll\min\{4d,8\left(  J-d\right)  ,4(J-2d)\}.
\end{equation}
From this consideration we observe that close to $d=0,J/2$ and $J$, the
perturbative approach (based on $|\Box\rangle$ and $|\times\rangle$) breaks
down and we should stay away from these points.

\begin{figure}[ptb]
\centering
{\includegraphics[width=8 cm]{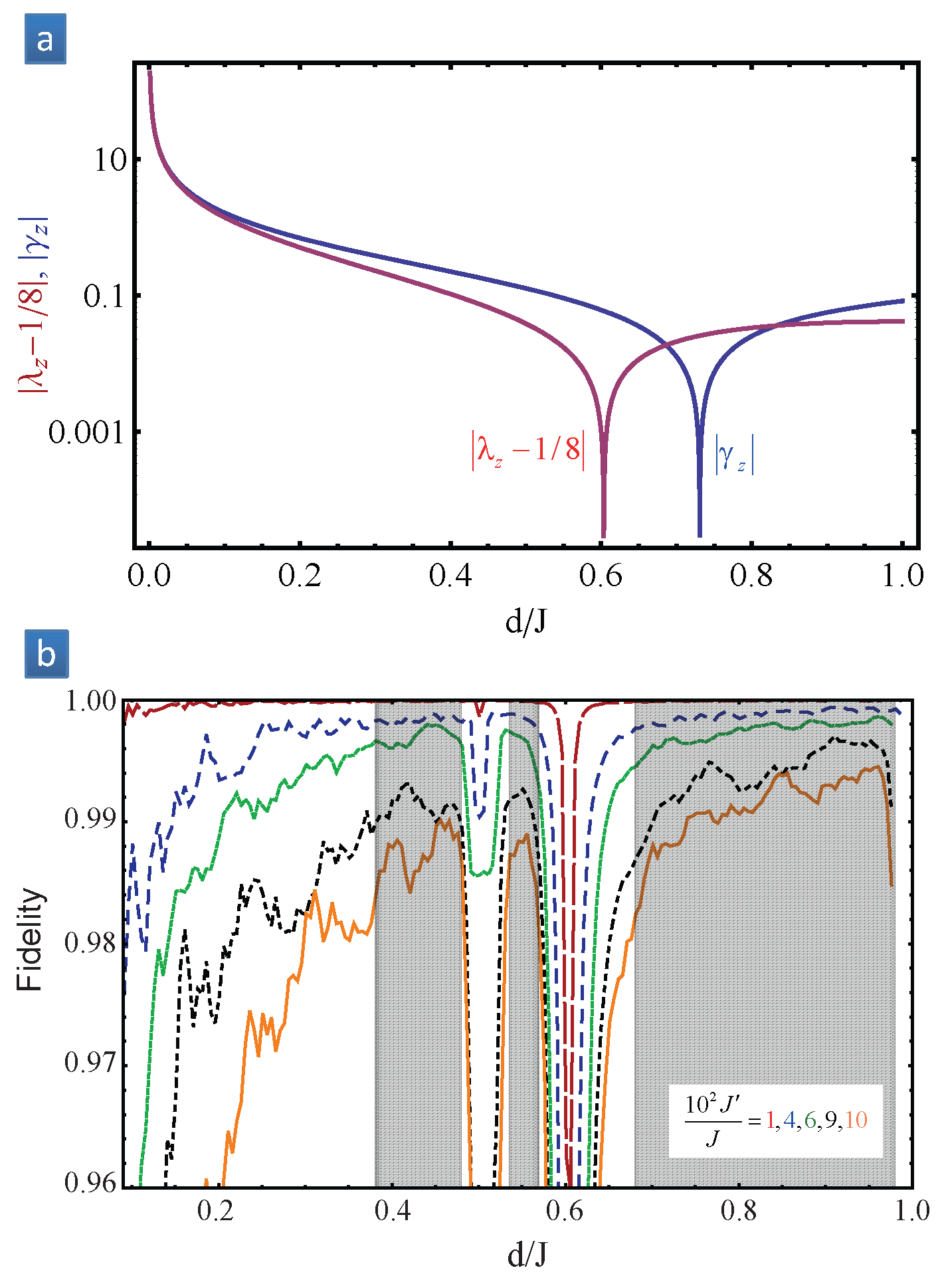}}
\caption[fig:PertFid]{(Color online) (a) Parameters of the effective
Hamiltonian for the general case $d\neq J$. At the point $d\approx0.62J$ , the
Ising term in the effective Hamiltonian vanishes, $\lambda_{z}=\frac{1}{8}$
(see text). (b) The controlled-phase gate fidelity $F$ on a superplaquette as
a function of $d/J$ for different rations of $J^{\prime}/J$. For same $d/J$,
the smaller $J^{\prime}/J$ the higher the fidelity. There are $4$ critical
points at which the fidelity drops considerably, these are $d/J\approx
0,0.5,0.62,1$. At $d/J\approx0$ and $0.5$, one of the singlet states becomes
degenerate with one $S=1$ state and consequently the effective Hamiltonian
breaks down. At the $d/J\approx0.62$ the Ising term vanishes and at $d=J$ the
rotating wave approximation used in the simplification of the effective
Hamiltonian becomes invalid. The shadow regions are the ones where the
achievable fidelity is higher than $0.98$.}%
\label{fig:PertFid}%
\end{figure}

As detailed in Appendix~\ref{app:d&J}, we can obtain the effective Hamiltonian%

\begin{align}
&  H_{i,i+1}^{eff}=(\frac{\Delta E}{2}-\frac{J^{^{\prime}2}\gamma_{z}}{J}%
)\sum_{j=i,i+1}\hat{\sigma}_{j}^{z}\label{eq:Heff}\\
&  -\frac{J^{^{\prime}2}}{J}\left[  \frac{1}{8}\hat{\vec{\sigma}}_{i}\cdot
\hat{\vec{\sigma}}_{i+1}+(\lambda_{z}-\frac{1}{8})\hat{\sigma}_{i}^{z}%
\hat{\sigma}_{i+1}^{z}\right]  ,\nonumber
\end{align}
where $\hat{\sigma}$ are effective Pauli matrices acting on the $|\Box
\rangle,|\times\rangle$ states, and
\begin{align}
\lambda_{z}  &  =\frac{1}{48}\left(  \frac{9J}{d}-\frac{8J}{d-3J}+2-\frac
{24J}{d+J}+\frac{J}{2J-d}\right) \\
\gamma_{z}  &  =\frac{1}{48}\left(  \frac{9J}{d}+\frac{8J}{d-3J}-8-\frac
{J}{2J-d}\right)  .
\end{align}
In Fig.~\ref{fig:PertFid}(a) the parameters $\lambda_{z}$ and $\gamma_{z}$ are
plotted as a function of $d/J$.

Within a superplaquette the term $\hat{\vec{\sigma}}_{i}\cdot\hat{\vec{\sigma
}}_{i+1}$ commutes with $H_{i,i+1}^{eff}$ (it only introduces a phase
$\phi_{T}=\frac{J^{^{\prime}2}}{\hbar J8}t_{c}$ in the effective triplet
subspace: $|\Box,\Box\rangle$ ,$|\times,\times\rangle$ and $(|\Box
,\times\rangle+|\times,\Box\rangle)/\sqrt{2}$ and $\phi_{S}=-\frac
{3J^{^{\prime}2}}{\hbar J8}t_{c}$ for the effective singlet states:
$(|\Box,\times\rangle-|\times,\Box\rangle)/\sqrt{2}$). Here $t_{c}$ stands for
the duration of the controlled-phase gate, i.e., $\frac{J^{^{\prime}2}}%
{J}t_{c}(\lambda_{z}-\frac{1}{8})=\hbar(2n-1)\pi/4$, with an integer
$n=1,2,\cdots$. Consequently $H_{i,i+1}^{eff}$ can be used to perform a
controlled-phase gate within a superplaquette.
We use the standard echo technique (i.e., $\pi$ pulses at $t_{c}/2$ and
$t_{c}$ for each of the encoded qubits) to remove the unwanted $\hat{\sigma
}_{j}^{z}$ term from the effective Hamiltonian in Eq.~(\ref{eq:Heff}). The
controlled-phase gate can be achieved by the unitary evolution%
\begin{equation}
U=X~e^{-i\left(  H_{intra}+H_{c}\right)  t_{c}/2}~X~e^{-i\left(
H_{intra}+H_{c}\right)  t_{c}/2},
\end{equation}
where $X$ represents the echo $\pi$ pules for the encoded qubits, which can be
achieved via intra-plaquette super-exchange couplings.

We use the exact diagonalization to calculate the controlled-phase gate
fidelity $F=\left\vert f\right\vert ^{2}$ with%
\begin{equation}
f=\frac{1}{N}\operatorname*{Tr}\left[  U_{c-phase}^{\dag}P~U~P\right]  ,
\end{equation}
where $P$ is the projection operator to the singlet subspace for each
plaquette. In Fig.~\ref{fig:PertFid}(b), we plot $F$ as a function of $d/J$
for different rations of $J^{\prime}/J$, with $n=1$. The Figure shows that at
the points $d\approx0.5J$, $d\approx0.62J$, $d=0$, $d=J$ there is an abrupt
drop of the fidelity, as the ratio $J^{\prime}/J$ is increased. The drop at
these points is expected since at $d\approx0.5J$ and $d\approx0$ one of the
singlet states becomes degenerate with one $S=1$ and consequently the
effective Hamiltonian breaks down. At $d\approx0.62J$ , $\lambda_{z}=1/8$ [see
Fig.~\ref{fig:PertFid}(a)] the Ising term vanishes and at $d=J$ the two
singlets become degenerate and the rotating wave approximation (assumed for
the derivation of the effective hamiltonian in Appendix~\ref{app:d&J}) is not
longer justified.

Away from these points the derived effective Hamiltonian provides a good
description of the dynamics and for values of $J^{\prime}\approx0.1J$ one can
get a gate fidelity above $0.98$. In the plot we highlight with a gray shadow
the $d/J$ parameter regime where the achievable fidelity is above $0.98$.
However among these shadow regions only the regime $d/J<0.5$ is experimentally
achievable using the lattice geometry described early in this section. The
small fluctuations in the fidelity curves are due to the non-energy preserving
terms neglected to obtain the effective hamiltonian (see
Appendix~\ref{app:d&J}), which can be suppressed when $J^{\prime}\ll
\sqrt{8\left(  J-d\right)  J}$.

\begin{figure}[ptb]
\centering
\centering
{\includegraphics[width=8.7 cm]{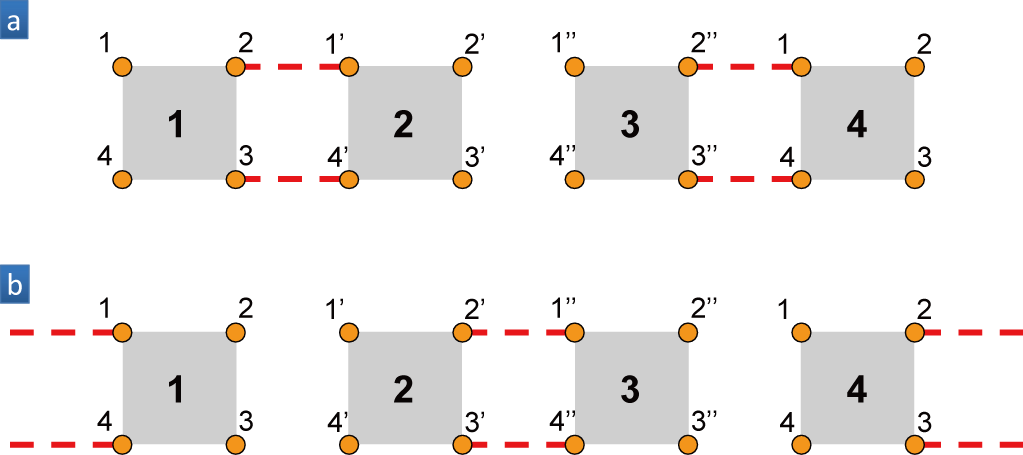}}%
\caption[fig:PertClusterState]{(Color online) For the $d<J$, the cluster state
generation has to be applied in two steps (a) and (b). }%
\label{fig:PertClusterState}%
\end{figure}

\begin{figure}[ptb]
\centering
{\includegraphics[width=6 cm]{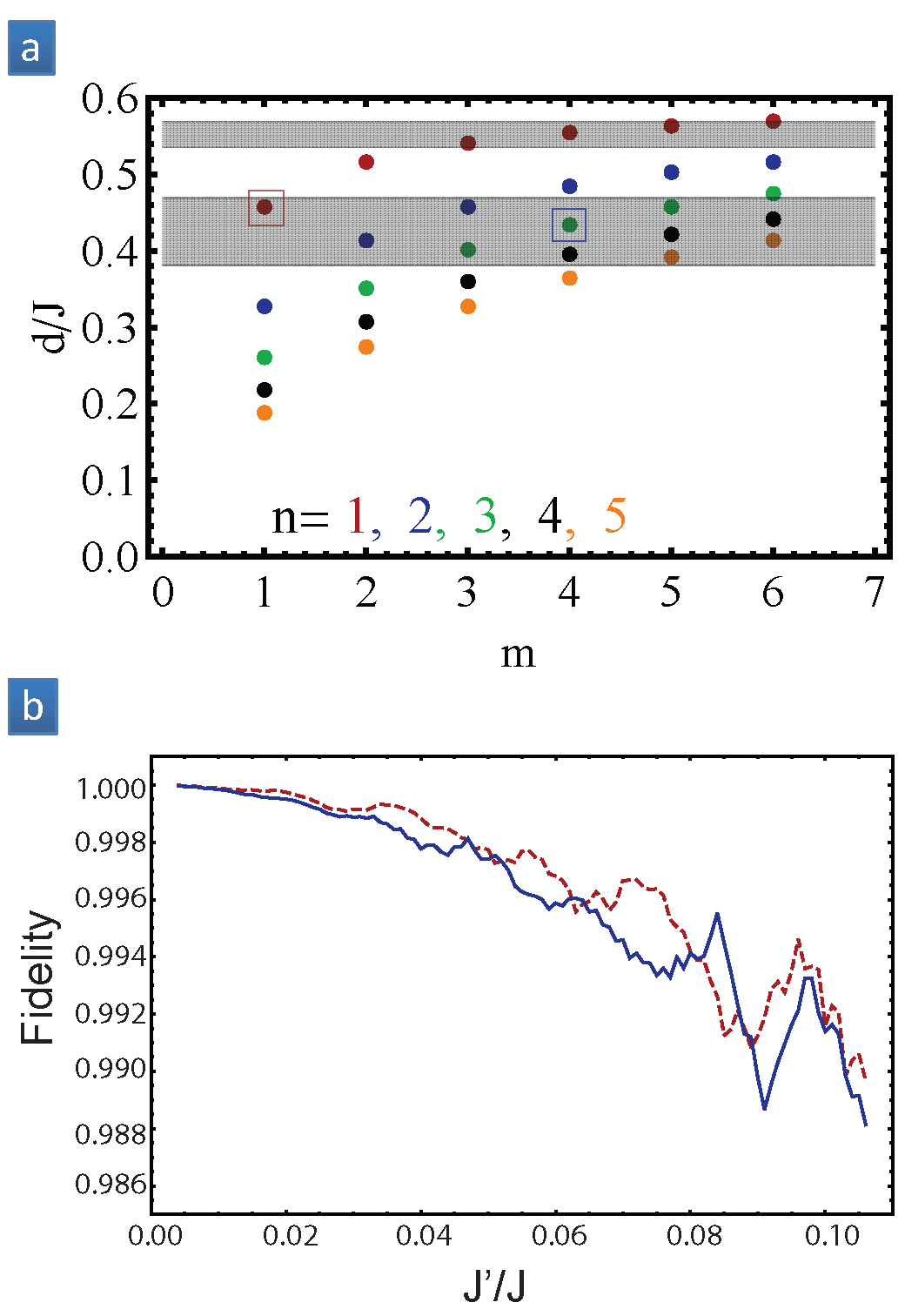}}
\caption[fig:PertExample]{(Color online) Examples of controlled-phase gate
with perturbative approach. (a) The values of $d/J$ that satisfy the
conditions stated in Eqs.~ (\ref{eq:con}). The shadow regions highlight the
regime where the phase gate fidelity can be larger than $0.98$ for $J^{\prime
}/J<0.1$. (b) The controlled-phase gate fidelity $F$ as a function of
$J^{\prime}/J$ can be calculated by exact numerical diagonalization. The red
dashed (or blue solid) curve is for $\left(  n,m\right)  =\left(  1,1\right)
$ (or $\left(  3,4\right)  $), which is also indicated by the red (blue)
square in panel (a).}%
\label{fig:PertExample}%
\end{figure}

The Heisenberg term $\hat{\vec{\sigma}}_{i}\cdot\hat{\vec{\sigma}}_{i+1}$, on
the other hand, does not commute with $H_{i+1,i+2}^{eff}$ and consequently the
phase gate can not be applied simultaneously to all plaquettes. Instead it has
to be applied first between the superplaquettes formed by the plaquettes
$2i+1,2i+2$ and subsequently between the superplaquettes formed from
plaquettes $2i+2,2i+3$ (see Fig.~\ref{fig:PertClusterState}). Additionally in
order to create a cluster state across all the plaquette array, it is required
to fine tune the parameters and time evolution to eliminate the different
phase accumulated by the triplet and singlet states in the encoded spin basis
due to the Heisenberg term at $t_{c}$. Consequently for multipartite
entanglement generation not all $d/J$ values are allowed but only the ones
which satisfy the following conditions:%

\begin{align}
\phi_{T}-\phi_{S}  &  =2\pi m\label{eq:con}\\
\frac{J^{^{\prime}2}}{J}t_{c}(\lambda_{z}-\frac{1}{8})  &  =\hbar
(2n-1)\frac{\pi}{4},
\end{align}
where $n$ and $m$ are integers. In Fig.~\ref{fig:PertExample}(a), we show a
set of allowed $d/J$ values which satisfy the conditions given by
Eq.~(\ref{eq:con}), for different $n$ and $m$ values. Here we also highlight
with a gray shadow the corresponding $d/J$ values which yield a fidelity
higher than 0.98 for $J^{\prime}/J<0.1$. In Fig.~\ref{fig:PertExample}(b) we
show two examples of traces of the phase gate fidelity vs $J^{\prime}%
/J$:$(n,m)=(1,1),(3,4)$ (indicated in panel a. by a square) computed by exact
diagonalization of the superplaquette Hamiltonian. The figure shows that it is
always possible to find parameters which allow for a high gate fidelity.
However, here we are only including errors due to higher order terms neglected
in the derivation of the effective Hamiltonian. In realistic experiments other
external errors such as lattice inhomogeneities are always present, which can
be minimized at the expense of a larger $J^{\prime}/J$ ratio (faster
evolution). There is consequently a trade off between faster time evolution
and small perturbative corrections.

In conclusion we have presented a scheme to perform controlled-phase gates in
the encoded singlet subspace. This perturbative scheme has two advantages: (1)
the plaquettes are always in the decoherence free subspace, (2) it is easy to
implement as it only relays on the coherent dynamical evolution without
further manipulations. Due to the fact that the dynamics is determined by a
second order effective Hamiltonian the achievable fidelity with the proposed
schemes can become very high but at the cost of slower time evolution. If the
strongly interacting regime is reached by using a Feshbach resonance, one can
achieve values of $J$ of order of $100$ Hz and therefore cluster generation
times of order $0.1-1$ sec. These generation times are slow but longer than
the encoded qubits decoherence time due to their insensitivity against
environmental decoherence \cite{Trotzky08}.

\subsection{Optimal control approach\label{sec:OptimalControlApproach}}

We now consider the optimal control approach to fast, high fidelity
implementation of the controlled-phase gate [between the horizontal
neighboring plaquettes $\left(  1,2,3,4\right)  $ and $\left(  1^{\prime
},2^{\prime},3^{\prime},4^{\prime}\right)  $ as shown in
Fig.~\ref{fig:GeoPhaseGate4Boson1}(a)]. The key challenge here is to identify
an efficient set of operators that
(i) enable the unitary evolution of the controlled-phase gate, and (ii) are
feasible using optical superlattices as well. We first provide a set of
operators sufficient to achieve the controlled-phase gate with arbitrary
precision. After that we numerically find the pulse sequences for these
operators to implement the controlled-phase gate.

\subsubsection{Choice of operators}

In principle, tunable Heisenberg superexchange interactions are sufficient for
the controlled-phase gate by coupling all eight sites \cite{Bacon00}. However,
for the optical lattice experiments, we would like to achieve the
controlled-phase gate by coupling as few sites as possible, preferably using
global rotations for all spins and Heisenberg/Ising interactions between
neighboring sites \cite{Trotzky08,Duan03}.

According to Eqs.~(\ref{eq:SigmaZ23},\ref{eq:SigmaZ41}), we need at least two
sites from each plaquette to determine the $\sigma_{z}$ operator. Since the
controlled-phase gate depends on both $\sigma_{z}$ operators from the
plaquettes, we should consider at least four sites to implement the
controlled-phase gate. It turns out that coupling the four middle sites
$\left(  2,3,4^{\prime},1^{\prime}\right)  $ is sufficient to achieve the
controlled-phase gate, which significantly reduces the complexity compared
with the earlier proposal that couples all eight sites \cite{Bacon00}.

We consider the Hamiltonian
\begin{equation}
H\left(  t\right)  =\sum_{k=1}^{5}\alpha_{k}\left(  t\right)  O_{k},
\label{eq:H(t)}%
\end{equation}
where $\left\{  \alpha_{k}\left(  t\right)  \right\}  $ are the time-dependent
control variables for the set of operators%
\begin{align}
O_{1}  &  =\vec{s}_{2}\cdot\vec{s}_{3}\\
O_{2}  &  =\vec{s}_{1}^{\prime}\cdot\vec{s}_{4}^{\prime}\\
O_{3}  &  =s_{2,z}s_{1,z}^{\prime}+s_{3,z}s_{4,z}^{\prime}\\
O_{4}  &  =s_{2,x}+s_{3,x}+s_{1,x}^{\prime}+s_{4,x}^{\prime}\\
O_{5}  &  =s_{2,y}+s_{3,y}+s_{1,y}^{\prime}+s_{4,y}^{\prime}.
\end{align}

To justify that $H\left(  t\right)  $ can implement the controlled-phase gate
\begin{equation}
U_{\text{c-phase}}=\exp\left[  -i\frac{\pi}{4}\left(  1-\sigma_{z}\right)
\left(  1-\sigma_{z}^{\prime}\right)  \right]  ,
\end{equation}
we show that $\left(  1-\sigma_{z}\right)  \left(  1-\sigma_{z}^{\prime
}\right)  $ belongs to the Lie algebra generated by $\left\{  O_{k}\right\}
_{k=1,\cdots,5}$. We start with these five operators as the \emph{available
set} (AS), and calculate the commutators among the AS operators. We then
expand the AS by adding new commutators that are not linear combinations of
the AS operators. We denote the number of linearly independent AS operators as
the dimension of the AS. We repeat the process of calculating the commutators
and expanding the AS, until its dimension does not increase any more. We use
Mathematica to iterate the process of expanding the AS until it saturates at
dimension $80$ (including the identity operator that commutes with all other
operators). Finally, we verify that $\left(  1-\sigma_{z}\right)  \left(
1-\sigma_{z}^{\prime}\right)  \sim\left(  \vec{s}_{2}\cdot\vec{s}_{3}\right)
\left(  \vec{s}_{1}^{\prime}\cdot\vec{s}_{4}^{\prime}\right)  $ is a linear
combination of the AS operators. Therefore, according to the local properties
from the Lie algebra the set of operators $\left\{  O_{k}\right\}
_{k=1,\cdots,5}$ is sufficient to implement the controlled-phase gate. The
remaining task is to find the solution for $\left\{  \alpha_{k}\left(
t\right)  \right\}  $.

\begin{table*}[tbp] \centering
\begin{tabular}
[c]{c|c|c|c}\hline\hline
&
\begin{tabular}
[c]{c}%
\textbf{Geometric Phase}\\
\textbf{Approach}%
\end{tabular}
&
\begin{tabular}
[c]{c}%
\textbf{Perturbative}\\
\textbf{Approach}%
\end{tabular}
&
\begin{tabular}
[c]{c}%
\textbf{Optimal Control}\\
\textbf{Approach}%
\end{tabular}
\\\hline
Time scale for controlled-Z (CZ) gate & $1/J$ & $J/J^{\prime2}$ & $1/J$\\
Duration out of DFS & $1/t$ & $0$ & $1/J$\\\hline
Systematic errors & $\left(  t/U\right)  ^{2}$ & $\left(  J^{\prime}/J\right)
^{2}$ & $0$\\
Inhomogeneity errors & $\left(  \delta t/t\right)  ^{2},\left(  \delta
\Delta/t\right)  ^{2}$ & $\left(  \delta J/J\right)  ^{2},\left(  \delta
J^{\prime}/J^{\prime}\right)  ^{2}$ & $\left(  \delta J/J\right)  ^{2}%
$\\\hline
Sites per CZ gate & $4$ & $8$ & $4$\\
Simultaneous coupling & Yes & Two steps & Yes\\\hline
Major interactions &
\begin{tabular}
[c]{c}%
Superexchange\\
Single particle tunneling
\end{tabular}
&
\begin{tabular}
[c]{c}%
Superexchange\\
\end{tabular}
&
\begin{tabular}
[c]{c}%
Superexchange\\
Ising interaction
\end{tabular}
\\\hline
Superlattice wavelengths & $\lambda,2\lambda,4\lambda$ & $\lambda
,2\lambda,4\lambda,4\lambda/3$ & $\lambda,2\lambda$\\\hline
Vibrational levels & Ground + excited & Ground & Ground\\\hline
Physical process & Clear & Clear & Hard to interpret\\\hline
Control complexity & Medium & Low & High\\\hline\hline
\end{tabular}
\caption{Comparison among three approaches.
See Sec.~\ref{sec:ComparingThreeApproaches} for discussions.}
\label{tab:Comparison}%
\end{table*}%

\subsubsection{Smooth pulses}

We use an algorithm which can be interpreted as a continuous version of the
gradient ascent pulse engineering (GRAPE) \cite{Khaneja05,Werschnik07}, though
it is developed in an independent way \cite{Romero-Isart07}. The algorithm
based on optimal quantum control is summarized in
Appendix~\ref{App:SmoothPulses}. Comparing with the GRAPE method it has the
advantage that we can find solutions with specific boundary conditions (e.g.
pulses start and end at zero) and in terms of smooth (finite slope) functions
of time as well.

More specifically, the particular form of the coefficients $\left\{
\alpha_{k}\left(  t\right)  \right\}  $ are chosen to be finite sums of
sinusoidal functions
\begin{equation}
\alpha_{k}(t,x_{k1},\ldots,x_{kL})=\sum_{l=1}^{L}x_{kl}\sin\left(  \frac{l\pi
t}{T}\right)  ,
\end{equation}
each of which depends on $L$ parameters $\{x_{kl}\}$ ($l=1,\ldots,L$). As
mentioned before, note that they fulfill the convenient property that
$\alpha_{k}(0)=\alpha_{k}(T)=0$ and that they have a finite slope. Hence, we
need to optimize $K\times L$ (with $K=5$ operators in our case) parameters
$x_{kl}$ which maximize the fidelity $F=|f|^{2}$, where
\begin{align}
f  &  =\frac{1}{N}\operatorname*{Tr}[U_{\text{c-phase}}^{\dagger
}U(T;x)]\label{eq:fid}\\
&  =\frac{1}{N}\sum_{n=1}^{N}\langle\psi_{n}|U_{\text{c-phase}}^{\dagger
}U(T;x)|\psi_{n}\rangle,\nonumber
\end{align}
for a particular subspace of $N$ states $\{|\psi_{n}\rangle\}$ ($n=1,\ldots
,N$) of dimension $d\geq N$.

\begin{figure}[b]
\begin{center}
\includegraphics[width=8.5cm]{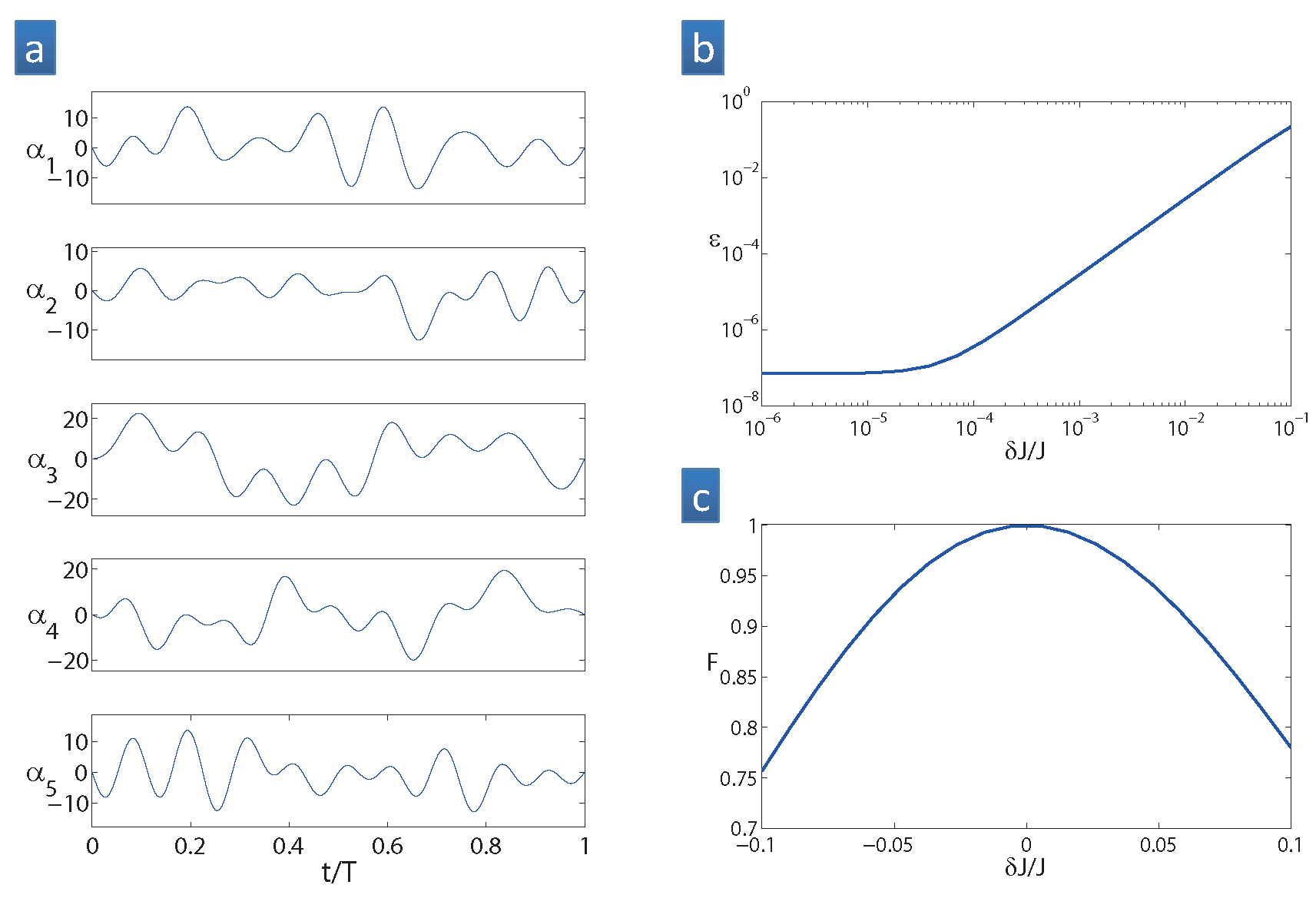}
\end{center}
\caption[fig:Smooth]{(Color online) (a) Smooth pulses with $L=20$ yielding to
infidelities smaller than $10^{-7}$. (b) Infidelity $\epsilon=1-F$ v.s. the
deviation $\delta J/J$ for the smooth pulses. The infidelity is constant at
$10^{-7}$ for $\delta J/J<10^{-4}$ and increases quadratically with $\delta
J/J$ for $\delta J/J>10^{-3}$. (c) Quadratic dependence of the fidelity with
$\delta J/J$.}%
\label{fig:Smooth}%
\end{figure}

In Fig.~\ref{fig:Smooth}(a) we show an example of pulses obtained with $L=20$,
which can attain very low infidelity $\varepsilon=1-F$, less than $10^{-7}$
(the value can be further reduced by improving the precision of the numerics).
In Fig.~\ref{fig:Smooth}(b) we plot the infidelity $\varepsilon$ as a function
of the \emph{relative deviation} $\delta J/J$. We assume that the couplings in
Eq.~(\ref{eq:H(t)}) are deviated from $\alpha_{k}$ to $\left(  1-\delta
J/J\right)  \alpha_{k}$; that is the system evolves under the deviated
Hamiltonian $(1-\delta J/J)H$. For simplicity, we consider the case that
$\delta J/J$ is time-independent. (For example, imperfect calibration of
barrier height or barrier thickness may induce such proportional,
time-independent deviation in superexchange couplings.) We find that the
infidelity remains a constant value (approximately $10^{-7}$) for very small
deviations (with $\delta J/J<10^{-4}$), while the infidelity scales as
$\left(  \delta J/J\right)  ^{2}$ for larger deviations (with $\delta
J/J>10^{-3}$) which is also plotted using the linear scale in
Fig.~\ref{fig:Smooth}(c). Such quadratic dependence to the deviation is not
uncommon, as the infidelity for single spin rotations also scales
quadratically with the deviation. The quadratic dependence can be regarded as
a direct consequence of the optimization procedure, which finds a local
minimum of the function with first order derivatives being zero.

\subsubsection{Experimental implementation}



We now briefly discuss the implementation of operators $\left\{
O_{k}\right\}  _{k=1,\cdots,5}$ (and $\left\{  O_{k}\right\}  $) for the
cluster state preparation. The operators of $O_{1}$ and $O_{2}$ can be
achieved by super-exchange interaction using superlattice techniques
\cite{Trotzky08}, while the operator of $O_{3}$ can be obtained from
spin-dependent tunneling in optical lattices \cite{Duan03}. Furthermore, we
note that the evolution of the Ising interactions $O_{3}$ between all
horizontal neighboring plaquettes can be performed simultaneously, because
they act on different groups of physical spins as illustrated in
Fig.~\ref{fig:ClusterStateMulti}. For the same reason, the operators of
$O_{4}$ and $O_{5}$ can be performed simultaneously for all spins by driving
the entire optical lattice with appropriate micro-wave pulses \cite{Mandel03}.
Therefore, the simultaneous controlled-phase gates between all horizontally
neighboring plaquettes can be achieved.

\subsection{Comparing three approaches\label{sec:ComparingThreeApproaches}}

We compare the three approaches (see Table~\ref{tab:Comparison}) in the
following aspects: (1) two relevant time scales: the time to implement the
controlled-phase gate, and the duration for the plaquette not being protected
by the DFS (which should be short compared to the coherence time outside the
DFS \footnote{%
The optimal control algorithm can also minimize the errors accumulated when
the system is outside the DFS during the optimal controlled evolution.%
}), (2) two types of errors contributing to the controlled-Z (CZ) gate infidelity: the
systematic errors from the approximations used in our analysis, and the
inhomogeneity errors due to the fact that the couplings (e.g., $t$, $\Delta$,
$J$, and $J^{\prime}$) are not exactly the same for all plaquettes, (3) the
number of sites involved for each controlled-phase gate: if each gate only
couples $4$ sites, the controlled-phase gates between all horizontal (or
vertical) neighbors can be achieved simultaneously; otherwise two sequential
steps are needed, (4) the major interactions, such as singlet particle
tunneling \cite{Folling07}, (Heisenberg) superexchange coupling
\cite{Trotzky08}, and Ising interaction \cite{Duan03}, (5) the wavelength
components needed to construct the superlattices, (6) the relevant vibrational
levels, (7) the interpretation of the physical process, and (8) the control
complexity for time-dependent parameters.

The maximum achievable fidelity for the CZ gate is limited by the systematic
and inhomogeneity errors. For the geometric phase approach, the systematic
error $\left(  t/U\right)  ^{2}$ is due to the off-resonant tunneling, which
is analyzed in Table~\ref{tab:EnergyDiff4Boson} and Table~\ref{tab:EnergyDiff}
for bosonic and fermonic particles, respectively; such off-resonant tunneling
can be suppressed by using Feshbach resonances to increase $U$ while keeping
the same tunneling rate $t$. For the perturbative approach, the
controlled-phase gate fidelity is $F>0.98$ for $J^{\prime}/J<0.1$
(Fig.~\ref{fig:PertExample}). For the optimal control approach, the systematic
error is only limited by the precision of the numerics [Fig.~\ref{fig:Smooth}%
(b,c)]. We note that it is important to suppress the inhomogeneity errors, as
all three approaches are sensitive to such imperfections. Replacing the
parabolic trap with the flat-bottom trap \cite{Gershnabel04} can be one
possible solution to reduce the inhomogeneity errors. It would also be
interesting to consider other approaches that are insensitive to the
inhomogeneity errors.

Overall, the geometric phase approach has the advantage of fast operational
time, short unprotected duration, and compatibility of simultaneous coupling.
The perturbative approach has the advantage of always being protected by the
DFS and favorable control complexity. The optimal control approach has the
advantage of fast operational time, vanishing systematic errors, and
compatibility of simultaneous coupling.

\section{Conclusion\label{sec:Conclusion}}

In conclusion, we have discussed preparation of large cluster states for
neutral atoms in optical superlattices. Each logical qubit is encoded in the
decoherence-free singlet subspace of four spins from the $2\times2$ plaquette,
so that it is insensitive to uniform magnetic field fluctuations along an
arbitrary direction. Besides arbitrary rotations of single logical qubit
achieved by superexchange interaction, we provide three different approaches
to couple the logical qubits from neighboring plaquettes, with their
properties summarized in Table~\ref{tab:Comparison}. These approaches may also
be applied to other quantum systems, such as quantum dots or Josephson
junction arrays.

We would thank Hans Briegel, Ignacio Cirac, Eugene Demler, Wolfgang D\"{u}r,
G\'{e}za Giedke, Vladimir Gritsev, and Bel\'{e}n Paredes for stimulating
discussions. AMR acknowledges support from the NSF (CAREER programs, ITAMP
grant). ORI, JJGR, and AS acknowledge financial support from the European
Commission (Integrated Project SCALA), from the Spanish M.E.C. (FIS2005-03619,
AP2005-0595, Consolider Ingenio2010 CSD2006-00019 QOIT, FIS2006-04885,
CAM-UCM/910758), from the Ramony Cajal Program, and from the Catalan
Government (SGR-00185).

\appendix{}

\section{Geometric Phase Approach\label{app:GeoPhaseApproach}}

\subsection{Geometric phase approach with bosonic particles}

We now justify the claim that the geometric phase $\pi$ is obtained for
$S_{2,3}\otimes T_{1^{\prime},4^{\prime}}$, while a trivial geometric phase
$0$ (or $2\pi$) is obtained for the other three cases ($S_{2,3}\otimes
S_{1^{\prime},4^{\prime}}$, $T_{2,3}\otimes S_{1^{\prime},4^{\prime}}$, and
$T_{2,3}\otimes T_{1^{\prime},4^{\prime}}$). This evolution implements the
controlled-phase gate up to a bit-flip of the logical qubit from the left plaquette.

We start by generalizing the on-site interaction Hamiltonian for site $i$ that
governs both the ground and excited vibrational levels (denoted as $a$ and $b$
respectively)%
\begin{align}
H_{i}  &  =\mu_{i}n_{i}+\omega_{i}n_{i}^{b}+\frac{1}{2}U_{i}^{aa}n_{i}%
^{a}\left(  n_{i}^{a}-1\right)  +\frac{1}{2}U_{i}^{bb}n_{i}^{b}\left(
n_{i}^{b}-1\right) \label{eq:HamBoson1}\\
&  +U_{i}^{ab}\left(  n_{i}^{a}n_{i}^{b}+\sum_{\sigma,\sigma^{\prime}%
}a_{i,\sigma}^{\dag}b_{i,\sigma^{\prime}}^{\dag}b_{i,\sigma}a_{i,\sigma
^{\prime}}+\sum_{\sigma,\sigma^{\prime}}b_{i,\sigma}^{\dag}b_{i,\sigma
^{\prime}}^{\dag}a_{i,\sigma}a_{i,\sigma^{\prime}}\right)  ,\nonumber
\end{align}
where $\mu_{i}$ is the energy off-set, $\omega_{i}$ is the vibrational
frequency, $U_{i}^{\alpha\beta}$ is the on-site interaction strength between
levels $\alpha$ and $\beta$ for site $i$. The particle number operators are
$n_{i}^{a}=\sum_{\sigma}a_{i,\sigma}^{\dag}a_{i,\sigma}$, $n_{i}^{b}%
=\sum_{\sigma}b_{i,\sigma}^{\dag}b_{i,\sigma}$, and $n_{i}=n_{i}^{a}+n_{i}%
^{b}$. Given large vibrational frequency $\omega_{i}\gg U_{i}^{ab}$, we may
safely neglect those energy non-conserving terms $\sum_{\sigma,\sigma^{\prime
}}b_{i,\sigma}^{\dag}b_{i,\sigma^{\prime}}^{\dag}a_{i,\sigma}a_{i,\sigma
^{\prime}}$.

\begin{figure}[t]
\begin{center}
\includegraphics[
width=6 cm
]{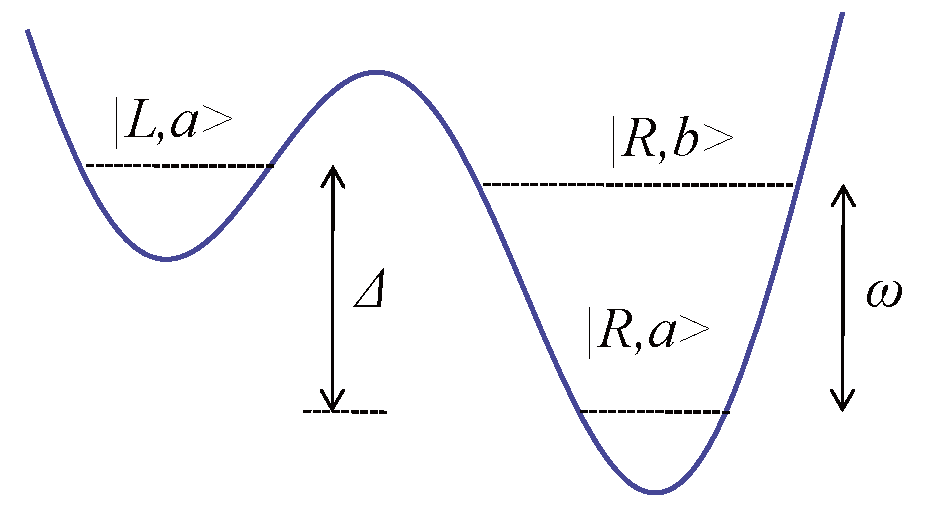}
\end{center}
\caption[fig:BiasedPotential]{(Color online) The biased potential for the
sites $L$ and $R$. The vibrational ground levels are $\left\vert
L,a\right\rangle $ and $\left\vert R,a\right\rangle $, with energy difference
$\Delta$. The vibrational excited level for the right site is $\left\vert
R,b\right\rangle $ with excitation energy $\omega$.}%
\label{fig:BiasedPotential}%
\end{figure}

For the biased potential between the two horizontal sites ($L$ and $R$) (as
shown in Fig.~\ref{fig:BiasedPotential}), we consider one vibrational level
for the left site and two levels for the right site:%
\begin{equation}
H_{L}=\Delta~n_{L}+U_{L}^{aa}n_{L}\left(  n_{L}-1\right)
\end{equation}%
\begin{align}
H_{R}  &  =\omega_{R}n_{R}^{b}+\frac{1}{2}U_{R}^{aa}n_{R}^{a}\left(  n_{R}%
^{a}-1\right)  +\frac{1}{2}U_{R}^{bb}n_{R}^{b}\left(  n_{R}^{b}-1\right)
\nonumber\\
&  +U_{R}^{ab}\left(  2n_{R}^{a}n_{R}^{b}+\vec{J}_{R}^{2}-\frac{n_{R}%
^{a}+n_{R}^{b}}{2}\left(  \frac{n_{R}^{a}+n_{R}^{b}}{2}+1\right)  \right)  ,
\label{eq:HamBoson2}%
\end{align}
where $\Delta=\mu_{L}-\mu_{R}$ is the bias in the potential (i.e., energy
difference between the ground levels for the two sites), and $\vec{J}_{R}$ is
the total spin for the right site (see Appendix~\ref{App:OnsiteInteraction}
for detailed derivation).

Given quantum numbers $\left(  n_{L},n_{R}^{a},n_{R}^{b},j_{R}\right)  $, the
on-site energies are%
\begin{equation}
E_{L}\left[  n_{L}\right]  =\Delta~n_{L}+U_{L}^{aa}n_{L}\left(  n_{L}%
-1\right)
\end{equation}%
\begin{align}
&  E_{R}\left[  n_{R}^{a},n_{R}^{b},j_{R}\right] \nonumber\\
&  =\omega_{R}n_{R}^{b}+\frac{1}{2}U_{R}^{aa}n_{R}^{a}\left(  n_{R}%
^{a}-1\right)  +\frac{1}{2}U_{R}^{bb}n_{R}^{b}\left(  n_{R}^{b}-1\right)
\nonumber\\
&  +U_{R}^{ab}f\left[  n_{R}^{a},n_{R}^{b},j_{R}\right]
\end{align}
where%
\begin{align}
&  f\left[  n_{R}^{a},n_{R}^{b},j_{R}\right] \nonumber\\
&  =2n_{R}^{a}n_{R}^{b}-\frac{n_{R}^{a}+n_{R}^{b}}{2}\left(  \frac{n_{R}%
^{a}+n_{R}^{b}}{2}+1\right)  +j_{R}\left(  j_{R}+1\right)
\end{align}
for $\frac{n_{R}^{a}+n_{R}^{b}}{2}\geq j_{R}\geq\frac{n_{R}^{a}-n_{R}^{b}}{2}%
$. Note that $n_{R}^{b}=0$ implies $f\left[  n_{R}^{a},0,j_{R}=n_{R}%
^{a}/2\right]  =0$. For $n_{R}^{b}=1$, we have $f\left[  n_{R}^{a}%
,1,j_{R}=\frac{n_{R}^{a}+1}{2}\right]  =2n_{R}^{a}$ and $f\left[  n_{R}%
^{a},1,j_{R}=\frac{n_{R}^{a}-1}{2}\right]  =n_{R}^{a}-1$.\footnote{The energy
difference for the co-tunneling from the left to the right site is $\delta
E_{2}\left[  n_{L}=2,n_{R}^{a},j_{R}\right]  =E_{L}\left[  2\right]
+E_{R}\left[  n_{R}^{a},0\right]  -E_{L}\left[  0\right]  -E_{R}\left[
n_{R}^{a},2\right]  =2\left(  \Delta-\omega\right)  +\left(  U_{L}^{aa}%
-U_{R}^{bb}\right)  -U_{R}^{ab}f\left[  n_{R}^{a},2,j_{R}\right]  ,$with
$f\left[  n_{R}^{a},2,j_{R}=\frac{n_{R}^{a}+2}{2}\right]  =4n_{R}^{a}$,
$f\left[  n_{R}^{a},2,j_{R}=\frac{n_{R}^{a}}{2}\right]  =3n_{R}^{a}-1$, and
$f\left[  n_{R}^{a},2,j_{R}=\frac{n_{R}^{a}-2}{2}\right]  =2n_{R}^{a}$.} Thus
the energy difference for the bosonic particle tunneling from the left site to
the right site is%
\begin{align}
&  \delta E_{1}\left[  n_{L},n_{R}^{a},j_{R}\right] \nonumber\\
&  =E_{L}\left[  n_{L}\right]  +E_{R}\left[  n_{R}^{a},0,n_{R}^{a}/2\right]
-E_{L}\left[  n_{L}-1\right]  -E_{R}\left[  n_{R}^{a},1,j_{R}\right]
\nonumber\\
&  =\left(  \Delta-\omega\right)  +U_{L}^{aa}\left(  n_{L}-1\right)
-U_{R}^{ab}f\left[  n_{R}^{a},1,j_{R}\right]  .
\end{align}
%

\begin{table}[tbp] \centering
\begin{tabular}
[c]{c|c|c}\hline\hline
$\left(  n_{L},n_{R}^{a}\right)  $ & $j_{R}=\left(  n_{R}^{a}-1\right)  /2$ &
$j_{R}=\left(  n_{R}^{a}+1\right)  /2$\\\hline
$\left(  1,0\right)  $ & -- & $\mathbf{0}$\\
$\left(  1,1\right)  $ & $0$ & $-2U_{R}^{ab}$\\
$\left(  1,2\right)  $ & $-U_{R}^{ab}$ & $-4U_{R}^{ab}$\\
$\left(  2,1\right)  $ & $2U_{L}^{aa}$ & $-2U_{R}^{ab}+2U_{L}^{aa}$\\
$\left(  2,2\right)  $ & $-U_{R}^{ab}+2U_{L}^{aa}$ & $-4U_{R}^{ab}+2U_{L}%
^{aa}$\\\hline\hline
\end{tabular}
\caption{
Energy difference associated with bosonic particle tunneling ($\delta E_{1}$)
for various initial number configurations $\left(  n_{L},n_{R}^{a}\right)  $
and the final total spin at the right site $j_{R}$. The bias is set to be
$\Delta=\omega$. The resonance condition $\delta E_{1}=0$ is fulfilled for
both cases: $\left(  n_{L},n_{R}^{a},j_{R}\right)  =\left(  1,0,1/2\right)  $
and $\left(  n_{L},n_{R}^{a},j_{R}\right)  =\left(  1,1,0\right)  .$
} \label{tab:EnergyDiff4Boson}%
\end{table}%

In Table~\ref{tab:EnergyDiff4Boson}, we list energy difference associated with
bosonic particle tunneling with $\Delta=\omega$. There are two possibilities
to fulfill the condition of resonant tunneling. The first case is $\left(
n_{L},n_{R}^{a},j_{R}\right)  =\left(  1,0,1/2\right)  $. This corresponds to
the resonant tunneling between the sites $\left(  2,1^{\prime}\right)  $ in
the highlighted upper right panel $S_{2,3}\otimes T_{1^{\prime},4^{\prime}}%
$\ of Fig.~\ref{fig:GeoPhaseGate4Boson2}(c), yielding a geometric phase $\pi$.
The second case is $\left(  n_{L},n_{R}^{a},j_{R}\right)  =\left(
1,1,0\right)  $. This corresponds to resonant tunneling for both pairs of
sites $\left(  2,1^{\prime}\right)  $ and $\left(  3,4^{\prime}\right)  $ as
shown in the left panel of Fig.~\ref{fig:GeoPhaseGate4Boson2}(c), yielding a
trivial geometric phase $2\pi$. The remaining cases are off-resonant
$\left\vert \delta E_{1}\right\vert \gg t$, which yields a trivial geometric
phase $0$. Therefore, we obtain a non-trivial geometric phase $\pi$ only for
$S_{2,3}\otimes T_{1^{\prime},4^{\prime}}$.

\subsection{On-site Interaction for Bosonic
Particles\label{App:OnsiteInteraction}}

We now derive Eq.~(\ref{eq:HamBoson2}) from Eq.~(\ref{eq:HamBoson1}). The key
step is to simplify the exchange term $\sum_{\sigma,\sigma^{\prime}%
}a_{i,\sigma}^{\dag}b_{i,\sigma^{\prime}}^{\dag}b_{i,\sigma}a_{i,\sigma
^{\prime}}$ using the particle number operators and the total spin operator.
For clarify, we drop the sub-index $i$ (or $R$).

Use the Schwinger representation, we define the spin operator $\vec{J}^{a}$
for the ground vibrational level%
\begin{align}
J_{x}^{a}  &  =\frac{1}{2}\left(  a_{\uparrow}^{\dag}a_{\downarrow
}+a_{\downarrow}^{\dag}a_{\uparrow}\right) \\
J_{y}^{a}  &  =\frac{1}{2i}\left(  a_{\uparrow}^{\dag}a_{\downarrow
}-a_{\downarrow}^{\dag}a_{\uparrow}\right) \\
J_{z}^{a}  &  =\frac{1}{2}\left(  n_{\uparrow}^{a}-n_{\downarrow}^{a}\right)
\\
j^{a}  &  =\frac{1}{2}\left(  n_{\uparrow}^{a}+n_{\downarrow}^{a}\right)
=\frac{n^{a}}{2},
\end{align}
with $\left(  \vec{J}^{a}\right)  ^{2}=j^{a}\left(  j^{a}+1\right)  $. Similar
definition for $\vec{J}^{b}$ can be introduced for the excited vibrational
level. Thus the total spin is $\vec{J}=\vec{J}^{a}+\vec{J}^{b}$.

We may rewrite the exchange interaction as%
\begin{align}
&  \sum_{\sigma,\sigma^{\prime}}a_{i,\sigma}^{\dag}b_{i,\sigma^{\prime}}%
^{\dag}b_{i,\sigma}a_{i,\sigma^{\prime}}\\
&  =\sum_{\sigma}a_{i,\sigma}^{\dag}a_{i,\sigma}b_{i,\sigma}^{\dag}%
b_{i,\sigma}+\sum_{\sigma}a_{i,\sigma}^{\dag}a_{i,\bar{\sigma}}b_{i,\bar
{\sigma}}^{\dag}b_{i,\sigma}\\
&  =\left(  2j^{a}j^{b}+2J_{z}^{a}J_{z}^{b}\right)  +2\left(  J_{x}^{a}%
J_{x}^{b}+J_{y}^{a}J_{y}^{b}\right) \\
&  =4j^{a}j^{b}+\vec{J}^{2}-\left(  j^{a}+j^{b}\right)  \left(  j^{a}%
+j^{b}+1\right) \\
&  =n^{a}n^{b}+\vec{J}^{2}-\frac{n^{a}+n^{b}}{2}\left(  \frac{n^{a}+n^{b}}%
{2}+1\right)  .
\end{align}
Plugging the above expression into Eq.~(\ref{eq:HamBoson1}) gives us
Eq.~(\ref{eq:HamBoson2}).

\subsection{Geometric phase approach with fermionic particles}

The procedure for the fermionic particles is almost the same as that for the
bosonic particles, except for the following three differences. First, the bias
of the energy off-set needs to be $\Delta=\omega+U_{R}^{ab}$ for fermionic
particles (whereas $\Delta=\omega$ for bosonic particles). Second, the
geometric phase $\pi$ is obtained from the resonant tunneling associated the
subspace $T_{2,3}\otimes S_{1^{\prime},4^{\prime}}$ for fermionic particles
(whereas it is associated with $S_{2,3}\otimes T_{1^{\prime},4^{\prime}}$ for
bosonic particles). Third, the geometric phase is $0$ for the remaining cases
for fermionic particles (whereas it might be either $0$ or $2\pi$ for bosonic particles).

For fermionic particles, the on-site interaction Hamiltonian for site $i$ that
governs both the ground and excited vibrational levels ($a$ and $b$) is%
\begin{align}
H_{i}  &  =\mu_{i}n_{i}+\omega_{i}n_{i}^{b}+U_{i}^{aa}n_{i,\uparrow}%
^{a}n_{i,\downarrow}^{a}+U_{i}^{bb}n_{i,\uparrow}^{b}n_{i,\downarrow}^{b}\\
&  +U_{i}^{ab}\left(  n_{i}^{a}n_{i}^{b}-\sum_{\sigma,\sigma^{\prime}%
}a_{i,\sigma}^{\dag}b_{i,\sigma^{\prime}}^{\dag}b_{i,\sigma}a_{i,\sigma
^{\prime}}\right)  ,\nonumber
\end{align}
where we have safely neglected the energy non-conserving terms $\sum
_{\sigma,\sigma^{\prime}}b_{i,\sigma}^{\dag}b_{i,\sigma^{\prime}}^{\dag
}a_{i,\sigma}a_{i,\sigma^{\prime}}$ for $\omega_{i}\gg U_{i}^{ab}$.

For the biased potential between the two horizontal sites ($L$ and $R$) as
shown in Fig.~\ref{fig:BiasedPotential}, we have%
\begin{equation}
H_{L}=\Delta~n_{L}+U_{L}^{aa}n_{L,\uparrow}n_{L,\downarrow},
\end{equation}%
\begin{align}
H_{R}  &  =\omega n_{R}^{b}+U_{R}^{aa}n_{R,\uparrow}^{a}n_{R,\downarrow}%
^{a}+U_{R}^{bb}n_{R,\uparrow}^{b}n_{R,\downarrow}^{b}\\
&  +U_{R}^{ab}\left(  n_{R}^{a}n_{R}^{b}-\sum_{\sigma}n_{R,\sigma}%
^{a}n_{R,\sigma}^{b}-\sum_{\sigma}a_{\sigma}^{\dag}a_{\bar{\sigma}}%
b_{\bar{\sigma}}^{\dag}b_{\sigma}\right)  ,\nonumber
\end{align}

Given quantum numbers $\left(  n_{L},n_{R}^{a},n_{R}^{b},j_{R}\right)  $, we
obtain the on-site energy%
\begin{equation}
E_{L}\left[  n_{L}\right]  =\Delta~n_{L}+U_{L}^{aa}\delta_{n_{L},2},
\end{equation}%
\begin{align}
E_{R}\left[  n_{R}^{a},n_{R}^{b},j_{R}\right]   &  =\omega n_{R}^{b}%
+U_{R}^{aa}\delta_{n_{R}^{a},2}+U_{R}^{bb}\delta_{n_{R}^{b},2}\\
&  +\frac{1}{2}U_{R}^{ab}\left(  n_{R}^{a}n_{R}^{b}+\eta_{n_{R}^{a},n_{R}%
^{b},j_{R}}\right)  ,\nonumber
\end{align}
where
\begin{equation}
\eta_{n_{R}^{a},n_{R}^{b},j_{R}}=\left(  3-4j_{R}\right)  \delta_{n_{R}^{a}%
,1}\delta_{n_{R}^{b},1}%
\end{equation}
for the spin dependent interaction. If $n_{R}^{a}=n_{R}^{b}=1$, $\eta=3$ for
spin singlet and $\eta=-1$ for spin triplet states; otherwise, $\eta=0$.

The energy difference associated with fermionic particle tunneling from the
left to the right site is%

\begin{align}
&  \delta E_{1}\left[  n_{L},n_{R}^{a},j_{R}\right] \nonumber\\
\equiv &  E_{L}\left[  n_{L}\right]  +E_{R}\left[  n_{R}^{a},0,\frac{1}%
{2}\delta_{n_{R}^{a},1}\right] \nonumber\\
&  -E_{L}\left[  n_{L}-1\right]  -E_{R}\left[  n_{R}^{a},1,j_{R}\right]
\nonumber\\
=  &  \left(  \Delta-\omega-\frac{1}{2}U_{R}^{ab}\left(  n_{R}^{a}+\eta
_{n_{R}^{a},1,j_{R}}\right)  \right)  +U_{L}^{aa}\delta_{n_{L},2},
\end{align}
By choosing $\Delta=\omega+U_{R}^{ab}$, we fulfill the resonance condition
$\delta E_{1}=0$ for $\left(  n_{L},n_{R}^{a}\right)  =\left(  1,2\right)  $.
In Table~\ref{tab:EnergyDiff}, we list the energy difference associated with
fermionic particle tunneling for various particle number configurations.%

\begin{table}[tbp] \centering
\begin{tabular}
[c]{c|c}\hline\hline
$\left(  n_{L},n_{R}^{a}\right)  $ & $\delta E_{1}$\\\hline
$\left(  1,0\right)  $ & $U_{R}^{ab}$\\
$\left(  1,1\right)  $ & $\mp U_{R}^{ab}$\\
$\left(  1,2\right)  $ & $\mathbf{0}$\\
$\left(  2,1\right)  $ & $-\frac{1}{2}U_{R}^{ab}+U_{L}^{aa}$\\
$\left(  2,2\right)  $ & $-U_{R}^{ab}+U_{L}^{aa}$\\\hline\hline
\end{tabular}
\caption{
Energy difference associated with fermionic particle tunneling ($\delta E_{1}$)
for different initial number configurations. The
bias is set to be $\Delta=\omega+U_{R}^{ab}$. For $\left(  n_{L},n_{R}^{a}\right)  =\left(  1,1\right)  $, the energy difference is $\mp U_{R}^{ab}$
for singlet and triplet states, respectively. The resonance condition $\delta
E_{1}=0$ is fulfilled only if $\left(  n_{L},n_{R}^{a}\right)  =\left(
1,2\right)  $.
} \label{tab:EnergyDiff}%
\end{table}%

Here are some remarks on the geometric phase approach. It is crucial to have
large and sufficiently different on-site interactions compared with the
tunneling rate $t$, because the virtual tunneling process may induce higher
order systematic errors $\sim\left(  t/U\right)  ^{2}$ in the accumulated
phase. One may use Feshbach resonances to enhance the on-site interaction and
suppress such errors. In addition, the tunneling rate $t$ and the energy
difference $\Delta$ should be as homogeneous as possible, since
inhomogeneities $\delta t$ and $\delta\Delta$ may induce leakage errors out of
the logical subspace with probability $\sim\left(  \frac{\delta t}{t}\right)
^{2}$ and $\left(  \frac{\delta\Delta}{t}\right)  ^{2}$, respectively. Optical
lattices in flat-bottom traps \cite{Gershnabel04} may efficiently eliminate
such inhomogeneity errors.

\section{Perturbative approach \label{app:d&J}}

\subsection{Effective Hamiltonian with $d=J$}

For $J=d$ (which might be achieved by placing an additional atom at the center
of the plaquette to block its occupancy due to interatomic repulsion
\cite{Buchler05}), the singlets are degenerated and isolated by an energy gap
$4J$ from the three-fold degenerate $S=1$ states. Since they are degenerate we
can choose any linear combinations of them as a basis. For convenience we
choose the states $|0\rangle$ and $|1\rangle$ as our basis. These two states
can be regarded as the two components of an effective pseudo-spin $1/2$ system.

In the absence of any coupling the singlet subspace of the chain is spanned by
product states of the effective pseudo-spin states $|0\rangle, |1\rangle$ at
the plaquettes and all four possible configurations are degenerate. A finite
$J^{\prime}$ breaks the degeneracy. Since the $H_{c}$ Hamiltonian does not
directly couple the effective pseudo spins, they get only coupled through
second order virtual processes to intermediate high energy states with $S>0$.
Denoting the left and right plaquettes as $i$ and $i+1$, the effective
Hamiltonian is given by:%

\begin{equation}
H^{eff}_{i, i+1}=\sum_{S,S^{\prime},q,q^{\prime}}\frac{ H_{c}|S^{q(S)}_{i}
S^{^{\prime}q^{\prime}(S^{\prime})}_{i+1}\rangle\langle S^{(q)}_{i}
S^{^{\prime}(q^{\prime})}_{i+1}| H_{c}}{2 E(0)-E(S)-E(S^{\prime})}%
\end{equation}
Here $q(S)$ labels the number of states within a plaquette with total spin
$S$. After some algebra one obtains that%

\begin{equation}
H_{i,i+1}^{eff}=-\frac{J^{^{\prime}2}}{3J}\left[  \tilde{\sigma}_{R}^{z}%
\tilde{\sigma}_{L}^{z}-\frac{1}{2}(\tilde{\sigma}_{R}^{z}+\tilde{\sigma}%
_{L}^{z})\right]  .
\end{equation}
where $\tilde{\sigma}_{R}^{z}$ are effective Pauli matrices acting on the
pseudo-spin states $|0\rangle,|1\rangle$. $H^{eff}$ is an effective Ising
hamiltonian and up to single logical qubit rotations it can be used to
generate cluster states encoded with the singlet basis. Since $H_{i,i+1}%
^{eff}$ commutes with $H_{i+1,i+2}^{eff}$, the cluster state generation can be
performed simultaneously in the 2D array of plaquettes.

\subsection{Effective Hamiltonian with $d<J$}

For $d<J$, we shall use the eigen-basis $|\Box\rangle$ and $|\times\rangle$.
We obtain the effective Hamiltonian using the second order perturbation theory:%

\begin{align}
&  H_{i,i+1}^{eff}=(\frac{\Delta E}{2}-\frac{J^{^{\prime}2}\gamma_{z}}{J}%
)\sum_{j=i,i+1}\hat{\sigma}_{j}^{z}\label{eq:eha}\\
&  -\frac{J^{^{\prime}2}}{J}\left[  \frac{1}{4}\hat{\sigma}_{i}^{x}\hat
{\sigma}_{i+1}^{x}+\lambda_{z}\hat{\sigma}_{i}^{z}\hat{\sigma}_{i+1}%
^{z}\right] \nonumber\\
&  -\frac{J^{^{\prime}2}}{J}\left[  -\frac{1}{4\sqrt{3}}(\hat{\sigma}_{i}%
^{x}\hat{\sigma}_{i+1}^{z}+\hat{\sigma}_{i}^{z}\hat{\sigma}_{i+1}^{x}%
)+\frac{1}{4\sqrt{3}}\sum_{j=i,i+1}\hat{\sigma}_{j}^{x}\right]  ,\nonumber
\end{align}
where $\hat{\sigma}$ are effective Pauli matrices acting on the $|\Box
\rangle,|\times\rangle$ states, and
\begin{align}
\lambda_{z}  &  =\frac{1}{48}\left(  \frac{9J}{d}-\frac{8J}{d-3J}+2-\frac
{24J}{d+J}+\frac{J}{2J-d}\right) \\
\gamma_{z}  &  =\frac{1}{48}\left(  \frac{9J}{d}+\frac{8J}{d-3J}-8-\frac
{J}{2J-d}\right)  .
\end{align}
In Fig.~\ref{fig:PertFid}(a) the parameters $\lambda_{z}$ and $\gamma_{z}$ are
plotted as a function of $d/J$.

The above Hamiltonian is more complex than the one derived for the previous
$J=d$ case, however if the inter-plaquette coupling is smaller than the energy
splitting between the two singlet states, $J^{\prime2}/J\ll\Delta E$, it is
energetically costly to flip an encoded spin and only the terms that preserve
the total effective magnetization are relevant. Consequently, in this regime
one can use an effective rotating wave approximation which consists on
neglecting the non-energy-preserving terms (last line in Eq.~\ref{eq:eha}). It
leads to a simpler effective Hamiltonian:%

\begin{align}
&  H_{i,i+1}^{eff}=(\frac{\Delta E}{2}-\frac{J^{^{\prime}2}\gamma_{z}}{J}%
)\sum_{j=i,i+1}\hat{\sigma}_{j}^{z}\\
&  -\frac{J^{^{\prime}2}}{J}\left[  \frac{1}{8}\hat{\vec{\sigma}}_{i}\cdot
\hat{\vec{\sigma}}_{i+1}+(\lambda_{z}-\frac{1}{8})\hat{\sigma}_{i}^{z}%
\hat{\sigma}_{i+1}^{z}\right] \nonumber
\end{align}

\section{Optimal quantum control with smooth pulses\label{App:SmoothPulses}}


Here we present the algorithm used to optimize the smooth functions
$\alpha_{k}$. This algorithm is significantly different from what is the
standard approach in optimal quantum control based on a Lagrangian formulation
\cite{Werschnik07}, and is an extension of the one used by some of the authors
in \cite{Romero-Isart07}.

Let us consider that we have the Hamiltonian
\begin{equation}
H(t;x)=\sum_{k=1}^{K}\alpha_{k}(t,x_{k1},\ldots,x_{kL})O_{k}=\sum_{k=1}%
^{K}O_{k}\sum_{l=1}^{L}x_{kl}J_{l}(t) \label{eq:HamJ}%
\end{equation}
where in our case $K=5$ and $J_{l}(t)=\sin(lt\pi/T)$. We have defined
$x\equiv\{x_{kl}\}$ as the set of $K\times L$ parameters that we wish to
optimize. The optimization is made such that the unitary evolution operator
$U(T;x)$
\begin{equation}
i\frac{d}{dt}U(t;x)=H(t;x)U(t;x) \label{schrod}%
\end{equation}
with initial condition $U(0;x)=\mathbb{I}$ gets as close as possible to a
desired unitary gate $U_{g}$ acting on a subspace $\{|\psi_{n}\rangle\}$ of
$N$ states of dimension $d\geq N$. This can be quantified with the fidelity
\[
f=\frac{1}{N}\mathrm{\operatorname*{Tr}}[U_{g}^{\dagger}U(T;x)]=\frac{1}%
{N}\sum_{n=1}^{N}\langle\psi_{n}|U_{g}^{\dagger}U(T;x)|\psi_{n}\rangle.
\]
To avoid complex numbers one can use either $F=\mathrm{Re}\{f\}$ or
$F=|f|^{2}$. For simplicity we will derive the algorithm using $F=\mathrm{Re}%
\{f\}$ (but we have presented the results with $F=|f|^{2}$ as this is not
sensible to irrelevant global phases). To maximize $F$ we need to compute the
derivative of $F$ with respect to the parameters $x$, i.e. $\partial
F/\partial x_{kl}$. If this can be done efficiently, then one can use any of
the multiple optimization algorithms which compute the optimal control. In
particular, the derivative and the formulas derived below are fed to Matlab's
nonlinear optimization toolbox \cite{Branch99}.

The partial derivatives of $F$ can be expressed
\[
\frac{\partial F}{\partial x_{kl}}=\frac{1}{N}\mathrm{Re}\sum_{n=1}^{N}%
\langle\psi_{n}|U_{g}^{\dagger}\frac{\partial}{\partial x_{kl}}U(T;x)|\psi
_{n}\rangle,
\]
which relates the gradient of $F$ to a derivative of the unitary operator
$U(T;x)$. Using second order perturbation theory (see the appendix in
\cite{Romero-Isart07} for details) the derivative of $U(T;x)$ can be expressed
as
\begin{equation}
\frac{\partial}{\partial x_{kl}}U(t;x)=-iU(t;x)\int_{0}^{t}d\tau
U(\tau;x)^{\dagger}\frac{\partial H(\tau;x)}{\partial x}U(\tau;x).
\label{derivative}%
\end{equation}
Hence, we get the following closed formula for the gradient of the fidelity
\begin{equation}
\frac{\partial F}{\partial x_{kl}}=\frac{1}{N}\mathrm{Im}\sum_{n=1}^{N}%
\int_{0}^{T}d\tau\langle\psi_{n}|U_{g}^{\dagger}U(T)U(\tau)^{\dagger}%
\frac{\partial H(\tau)}{\partial x_{kl}}U(\tau)|\psi_{n}\rangle.
\label{final-derivative}%
\end{equation}
(from hereafter we omit the $x$-dependence of $U$ and $H$ in order to ease the
notation). Though we have a closed formula, we still need to perform the
integral. To do so, we devise an efficient procedure which is based on solving
three sets of ordinary differential equations (ODEs). First note, that the
integral in Eq.~(\ref{final-derivative}) can be transformed into $N\times
K\times L$ ODEs%

\begin{equation}
\frac{d}{dt} \mathfrak{f}_{nkl}(t) = \frac{1}{N} \text{Im} \langle\psi_{n}|
U_{g}^{\dagger}U(T) U(t)^{\dagger}\frac{\partial H(t)}{\partial x_{kl}} U(t)
|\psi_{n}\rangle
\end{equation}
with initial conditions $\mathfrak{f}_{nkl}(0)=0$. Thus, we have that
\begin{equation}
\frac{\partial F}{\partial x_{kl}} =\sum_{n=1}^{N} \mathfrak{f}_{nkl}(T).
\end{equation}

Then, the algorithm to obtain the gradient of the fidelity, which is fed to
Matlab's nonlinear optimization toolbox \cite{Branch99}, is given by:

\begin{enumerate}
\item Solve the $N$ ODEs
\begin{equation}
i \frac{d}{dt} |\xi_{n}(t)\rangle= H(t) |\xi_{n}(t)\rangle, \label{reverse}%
\end{equation}
with initial condition $|\xi_{n}(T)\rangle= U_{g}|\psi_{n}\rangle$ and moving
backwards in time from $T$ to $t$.

\item Solve the $2N$ ODEs
\begin{subequations}
\begin{align}
i\frac{d}{dt}|\psi_{n}(t)\rangle &  :=H(t)|\psi_{n}(t)\rangle\\
i\frac{d}{dt}|\xi_{n}(t)\rangle &  :=H(t)|\xi_{n}(t)\rangle
\end{align}
with the initial conditions $|\psi_{n}(0)\rangle=|\psi_{n}\rangle$, and
$|\xi_{n}(0)\rangle=U(T)^{\dagger}U_{g}|\psi_{n}\rangle$ computed before.

\item Solve the $K\times L$ ODEs
\end{subequations}
\begin{equation}
\frac{d}{dt}\tilde{\mathfrak{f}}_{kl}=\frac{J_{l}(t)}{N}\sum_{n=1}%
^{N}\text{Im}\langle\xi_{n}(t)|O_{k}|\psi_{n}(t)\rangle
\end{equation}
with initial condition $\tilde{\mathfrak{f}}_{kl}(0)=0$. We have defined
$\tilde{\mathfrak{f}}_{kl}=\sum_{n}\mathfrak{f}_{nkl}$ and we have used the
expression of the particular Hamiltonian of Eq.~(\ref{eq:HamJ}). This step can
be done simultaneously with step 2 so that $|\psi_{n}(t)\rangle$ and $|\xi
_{n}(t)\rangle$ need not to be stored.

\item The derivatives of the fidelity are then given by
\begin{equation}
\frac{\partial F}{\partial x_{kl}}=\tilde{\mathfrak{f}}_{kl}(T).
\end{equation}

\end{enumerate}

\bibliographystyle{apsrev}
\bibliography{ref}

\end{document}